\documentclass[journal]{IEEEtran}
\usepackage[utf8]{inputenc}
\usepackage{amsmath}
\usepackage{amssymb}
\usepackage{amsfonts}
\usepackage{mathtools}
\usepackage{bm}
\usepackage{graphicx}
\usepackage{booktabs}
\usepackage{multirow}
\usepackage{array}
\usepackage{tabularx}
\usepackage{algorithm}
\usepackage{algpseudocode}
\usepackage{url}
\usepackage{tcolorbox}
\usepackage{cuted}
\usepackage{caption}
\usepackage{fvextra}
\usepackage{xcolor}
\usepackage{cite}
\usepackage{balance}
\usepackage[hidelinks]{hyperref}
\usepackage[normalem]{ulem}
\hypersetup{
  colorlinks=true,
  linkcolor=blue,
  citecolor=blue,
  urlcolor=blue,
  filecolor=blue
}

\begin{document}

\title{Self-Adaptive Multi-Agent LLM-Based Security Pattern Selection for IoT Systems}

\author{
Saeid~Jamshidi,
Foutse~Khomh,
Carol~Fung,
and Kawser~Wazed~Nafi

\thanks{
S. Jamshidi, F. Khomh, and Kawser Wazed Nafi are with the SWAT Laboratory,
Polytechnique Montréal, Montréal, QC, Canada
(e-mail: \{saeid.jamshidi, foutse.khomh, kawser.wazed-nafi\}@polymtl.ca).
}
\thanks{
C. Fung is with the Concordia Institute for Information Systems Engineering (CIISE),
Concordia University, Montréal, QC, Canada
(e-mail: \{carol.fung\}@concordia.ca).
}
}

\maketitle

\begin{abstract}
The adoption of Internet of Things (IoT) systems at the network edge of smart architectures is increasing rapidly, intensifying the need for security mechanisms that are both adaptive and resource-efficient. In such environments, runtime defence mechanisms are no longer limited to detection alone but become a resource-constrained task of selecting mitigation actions. Security controls must be carefully selected, combined, and executed under latency, energy, and computational constraints, while preventing unsafe interactions between controls. Existing approaches predominantly rely on static rule sets and learned policies, which provide limited guarantees of feasibility, conflict safety, and execution correctness in resource-constrained edge settings.  To address this limitation, we introduce \emph{ASPO}, a self-adaptive multi-agent security pattern selection that integrates Large Language Model (LLM)-based reasoning with deterministic enforcement within a MAPE-K control loop. ASPO explicitly separates stochastic decision generation from execution: LLM agents propose candidate mitigation portfolios, while a deterministic optimisation core enforces closed-world action integrity, conflict-free composition, and resource feasibility at every decision epoch. We deploy ASPO on a distributed edge-gateway testbed and evaluate it across two workloads, each comprising 500 and 1000 runtime security decisions, using replayed IoT attack traffic. In addition, the results demonstrate invariant safety properties, including 100\% conflict-free activation, consistent resource feasibility across workloads, and stable pattern dominance with perfect rank preservation. Importantly, deeper decision exploration reduces extreme-case execution costs, compressing tail latency and energy overheads by 21.9\% and 23.1\%, respectively, without increasing mean energy consumption.
\end{abstract}

\begin{IEEEkeywords}
IoT security, edge gateway, security, large language models, multi-agent LLMS.
\end{IEEEkeywords}

\section{Introduction}
\label{intro}
The rapid expansion of Internet of Things (IoT) systems has shifted security enforcement from centralised infrastructures to resource-constrained edge gateways \cite{zhukabayeva2025cybersecurity,hudda2025review,zhang2025towards}. These gateways aggregate traffic from heterogeneous devices and serve as the first layer of defence, operating under constraints on CPU, memory, latency, and energy \cite{trigka2025edge,sheikh2025survey}. As a result, runtime security at the edge is not merely a detection problem, but a constrained decision problem in which mitigation mechanisms must be selected and composed to suppress attacks while preserving system stability \cite{mahavaishnavi2025secure,batewela2025addressing}.\\
IoT security patterns provide reusable design solutions such as authentication, segmentation, and secure communication \cite{fernandez2021design,rajmohan2022decade}. However, they are largely static, and existing work does not address how to dynamically select, compose, and order patterns at runtime under resource constraints \cite{nguyen2023deep,wang2023reinforcement}. In practice, gateways must determine which mitigation actions to activate, whether they can coexist safely, and whether their combined cost remains feasible.\\
Adaptive approaches based on Deep Reinforcement Learning (DRL) learn mitigation policies from interaction \cite{lyu2025multi,kavianifar2025rs}. Our prior work \cite{jamshidi2025dynamic,jamshidi2025understanding} showed that DRL-based IDS can optimise decisions while accounting for latency and energy costs \cite{zhang2023energy}. However, DRL-based IDS remains a policy-learning paradigm that lacks explicit reasoning and cannot guarantee conflict-free, valid, and resource-feasible execution, since safety constraints are only implicitly encoded in the reward signal. Large Language Models (LLMs) offer reasoning capabilities and enable multi-agent decision-making workflows \cite{openai2023gpt4,huang2023agentsurvey,wu2023autogen,li2023camel,park2023generative}. However, LLM outputs are stochastic and may produce infeasible and unsafe mitigation plans without control \cite{greshake2023prompt,tran2025multi,rebedea2023nemoguardrails,jain2023baseline}. Consequently, existing approaches remain fragmented: DRL provides adaptation without guarantees, optimisation enforces constraints without contextual reasoning, and LLMs enable reasoning without safe execution. This limitation reflects a structural incompatibility between reasoning and enforcement. Empirical evidence from prior work \cite{jamshidi2025dynamic,jamshidi2025understanding} indicates that unconstrained mitigation selection often produces portfolios that violate conflict and resource feasibility requirements under dynamic edge conditions. This demonstrates that lightweight, unconstrained decision mechanisms are insufficient and underscores the need for an architecture that explicitly separates reasoning from deterministic enforcement. Therefore, the central research gap is the lack of a runtime mitigation mechanism that simultaneously supports context-aware decision-making, enforces the validity of closed-world actions, and guarantees conflict-free, resource-feasible execution.\\
To address these limitations, we introduce \emph{ASPO}, a self-adaptive multi-agent IoT security pattern selection grounded in the MAPE-K (Monitor, Analyse, Plan, Execute, Knowledge) autonomic control model \cite{zhou2019comprehensive,kephart2003autonomic}. The MAPE-K loop is required in this context to enforce a structured separation among monitoring, reasoning, and execution stages, enabling the controlled integration of stochastic LLM-based deliberation with deterministic enforcement, while ensuring that runtime decisions remain auditable, reproducible, and bounded by safe execution constraints. The need for such a solution arises from a limitation of existing approaches: runtime security decisions at the edge must simultaneously reason about complex threat contexts while ensuring that only safe, feasible, and executable actions are enforced under resource constraints. This requirement cannot be satisfied by existing approaches in isolation, since policy-learning methods provide adaptive behaviour without deterministic safety guarantees, while optimisation-based methods enforce feasibility without the ability to perform contextual reasoning. ASPO explicitly resolves this limitation by integrating context-aware reasoning with deterministic enforcement within a unified decision process. ASPO addresses this gap by operationalising design-time IoT security patterns as executable runtime primitives within a closed-world catalogue, including \emph{Security Logger/Auditor}, \emph{Secure Fog}, \emph{Security Segmentation}, \emph{Secure Distributed Publish/Subscribe}, \emph{Permission Control}, \emph{Whitelist}, \emph{Blacklist}, \emph{Outbound-Only Connection}, \emph{Personal Zone Hub}, and \emph{Trusted Communication Partner}. Each pattern is associated with explicit capability requirements, resource costs, and interaction constraints, enabling formal reasoning about safe composition at runtime. The system is self-adaptive because mitigation decisions are recomputed at each decision epoch based on updated telemetry, evolving threat context, and real-time resource headroom. Importantly, ASPO operates downstream of the detection layer and focuses on mitigation selection rather than attack detection; therefore, its objective is not classification accuracy but the correctness and feasibility of runtime decision-making under constraints. Conceptually, ASPO shifts runtime defence from policy learning to agentic deliberation with deterministic enforcement. Specialised LLM agents generate schema-constrained candidate portfolios conditioned on structured context, while a deterministic, conflict-aware optimisation core enforces closed-world action integrity, conflict-free composition, and resource feasibility before execution. Within ASPO, the \textit{Monitor} stage captures gateway telemetry and resource indicators, and the \textit{Analyse} stage constructs a structured threat representation including attack class, severity, confidence, behavioural evidence, and explicit operational budgets. The \textit{Plan} stage employs a multi-agent architecture comprising context, reasoning, constraint, planning, and auditing agents, whose outputs are bounded within the closed-world catalogue, and no agent directly executes mitigation commands. Final actuation is determined exclusively by a deterministic validation core, ensuring that all activated mitigation portfolios remain safe, feasible, and executable under latency and energy constraints. This separation establishes a verifiable decision boundary and enables traceable, auditable mitigation selection by explicitly validating constraints and preserving decision traces at each decision epoch. The main contributions of this work are summarised as follows:

\begin{itemize}
\item \textbf{From policy learning to safe decision-making: a hybrid agentic-deterministic paradigm.}
We introduce ASPO, a novel runtime security solution that moves beyond conventional policy learning by combining multi-agent LLM-based reasoning with a deterministic validation core. This design enables context-aware mitigation decisions while preventing unsafe, infeasible, and out-of-catalogue actions at execution time, thereby improving the reliability and explainability of runtime security decisions.
\item \textbf{Deterministic enforcement of security composition under real-world constraints.}
We reformulate runtime mitigation selection as a closed-world, conflict-aware optimisation problem in which security patterns are treated as executable primitives with explicit interaction rules, capability constraints, and resource costs. In contrast to prior approaches, ASPO guarantees conflict-free composition and adherence to latency, energy, and computational budgets as hard constraints at every decision epoch.
\item \textbf{Operational validation on a distributed edge testbed under realistic attack conditions.}
We implement ASPO on a distributed edge-gateway testbed and evaluate it using the Bot-IoT dataset across diverse attack scenarios. Extensive experiments demonstrate invariant safety enforcement, stable mitigation behaviour, and controlled latency and energy overheads under workload scaling, supporting practical deployment in resource-constrained IoT networks.
\end{itemize}

The remainder of this paper is organised as follows. Section~\ref{sec:related_work} reviews related work. Section~\ref{Threat Model and Security Assumptions} presents the threat model and system assumptions. Section~\ref{Methodological} describes the proposed ASPO methodology. Section~\ref{sec:experimental_setup} outlines the experimental setup. Section~\ref{sec:results} reports the experimental results. Section~\ref{Discussion} discusses the findings and implications. 
Section~\ref{Limitations and Future Work} highlights limitations and future directions. Section~\ref{Conclusion} concludes the paper.

\section{Related Work}
\label{sec:related_work}
This section presents the related work and reviews existing approaches relevant to runtime security mitigation in IoT networks.

\subsection{Dynamic Mitigation and Pattern Selection}
DRL has been widely explored as a mechanism for adaptive cyber defence, particularly in environments where attack behaviours and network conditions evolve over time. Recent work has applied RL to automated intrusion response, adaptive security policy tuning, and autonomous mitigation planning in cyber–physical and IoT networks \cite{iturbe2025rlresponse,suresh2025adaptive,saqib2025cloudrl}. These approaches enable systems to learn state-dependent mitigation strategies, where the state typically encodes observable system conditions such as traffic intensity, attack type, protocol behaviour, and resource utilisation levels, through interaction with the environment. Such formulations have been applied to intrusion response, traffic control, and adaptive policy tuning in IoT networks. More recently, multi-agent DRL has been investigated to distribute defensive decision-making across multiple learners, reflecting the decentralised and heterogeneous nature of modern cyber–physical systems \cite{wushishi2025d3o,ahmed2025distributed}. Hierarchical and cooperative MARL formulations \cite{mavroudis2025guidelines} have also been proposed to coordinate multiple defensive actions across network layers and hosts, improving scalability and adaptability in large cyber-defence environments. While RL/DRL methods offer strong adaptability, they primarily operate as policy learners and do not inherently provide structured deliberation mechanisms for composing multiple mitigations under explicit interaction constraints and hard resource budgets.

\subsection{LLM Agents for Cybersecurity Automation}
In parallel with learning-based control, LLMs have enabled a new class of agentic cybersecurity systems capable of reasoning over complex workflows. Recent work demonstrates that LLMs can interpret contextual signals, coordinate tools, and generate structured plans for security operations \cite{deng2024pentestgpt,shen2024pentestagent,xu2024autoattacker}. Beyond penetration-testing automation, more recent studies investigate LLM-driven autonomous security agents for monitoring, incident analysis, and defensive-reasoning pipelines \cite{xu2025forewarned,chowdhry2025evaluating}. These systems highlight LLMs' ability to support structured reasoning and workflow decomposition across complex cybersecurity tasks. However, their outputs remain stochastic and may result in infeasible actions unless execution is governed by deterministic validation mechanisms.

\subsection{Prompt Injection and the Need for Deterministic Actuation Boundaries}
The use of agentic LLM systems introduces additional risks, including prompt injection, tool misuse, and instruction manipulation via untrusted context. Recent research, therefore, explores defences against such attacks at inference time. DefensiveTokens proposes a lightweight mechanism that improves robustness against prompt injection by appending optimised special tokens without retraining the model \cite{chen2025defensivetokens}. Other recent works study safe tool use and constrained execution in LLM-driven systems, emphasising the importance of bounded action spaces and runtime verification to prevent unsafe autonomous decisions \cite{song2024secure,zhang2025allies}. While such techniques improve robustness, safe deployment in IoT gateways still requires architectural safeguards that ensure closed-world actuation, conflict-free mitigation composition, and strict feasibility within latency, energy, and computational budgets.

\subsection{Security Control Selection and Coordination Under Constraint}
Another group of researches formulates mitigation selection as a constrained planning and optimisation problem. These approaches emphasise the explicit modelling of operational trade-offs, thereby enabling security decisions to be audited and verified. Planning-based formulations have been applied to cost-aware security configuration, adaptive policy synthesis, and automated control selection under resource limits \cite{dutta2024security,ranaweera2025optimizing}. 
Recent work further explores multi-objective security optimisation in edge and IoT networks, where latency, computational overhead, and energy consumption must be jointly considered when selecting mitigation portfolios \cite{bisio2026distributed}. However, these approaches typically assume a predefined candidate set and do not integrate agentic reasoning layers that can interpret context and generate mitigation alternatives. They also rarely embed the process within an autonomic control loop while enforcing deterministic execution guarantees at each decision epoch.\\

The literature review indicates that existing research on adaptive IoT security evolves along three independent directions: RL-based mitigation selection, LLM-driven agentic cybersecurity automation, and constraint-aware security control optimisation. RL/DRL approaches provide dynamic adaptation but do not inherently enforce deterministic, conflict-free, and closed-world actuation guarantees at runtime. LLM-based multi-agent systems demonstrate reasoning and planning capabilities, yet they introduce stochastic behaviour and safety risks due to the absence of formally enforced boundaries. Optimisation-based control selection offers structured decision-making but lacks integration with agentic reasoning architectures. As a result, these research streams remain fragmented. To the best of this study’s knowledge, there is currently no unified solution that combines multi-agent LLM deliberation with deterministic, conflict-aware, and resource-feasible execution under explicit latency and energy constraints at the edge gateway level. ASPO addresses this gap through a hybrid agentic-deterministic architecture embedded within a formal MAPE-K control loop.

\section{Threat Model}
\label{Threat Model and Security Assumptions}
We consider an adversarial runtime environment in which the edge gateway receives telemetry, evidence tokens, and contextual inputs derived from network traffic that may be partially manipulated. The adversary’s objective is to impact runtime security decisions so that unsafe mitigation portfolios are activated, resource budgets are exceeded, and execution consistency is disrupted. The attacker therefore targets the decision process that maps structured context to security-pattern portfolios rather than attempting to compromise the detection stage itself. The attacker interacts with the system through observable traffic behaviour, including attack patterns such as DoS, DDoS, botnet activity, brute-force attempts, and port scanning, and indirectly through inputs consumed by the reasoning pipeline \cite{liu2023prompt,greshake2023not}. Three attack surfaces are considered: 1) Prompt-manipulation attempts aim to distort the semantic interpretation of telemetry by crafting adversarial traffic patterns that produce misleading evidence tokens, 2) Context poisoning attempts to bias structured context variables, such as threat severity, confidence, and resource indicators, through manipulated traffic characteristics, and 3) Tool-abuse attempts aim to push the decision pipeline toward infeasible and conflicting security patterns that could destabilize the gateway. Let the structured runtime context at epoch $t$ be $S_t$, constructed deterministically from telemetry. Adversarial impact is modelled as a bounded perturbation:
\begin{equation}
\tilde{S}_t = S_t + \delta_t, \qquad \|\delta_t\| \le \Delta ,
\end{equation}
where $\delta_t$ represents injected distortions in evidence tokens, resource signals, and contextual attributes derived from attack traffic. This perturbation may alter the stochastic candidate set proposed by the reasoning agents:
\begin{equation}
C_t = f_{\mathrm{LLM}}(\tilde{S}_t, \mathcal{P}),
\end{equation}
but it cannot directly determine the activated mitigation portfolio. Final activation is computed exclusively by the deterministic optimisation core:
\begin{equation}
Y_t^{\mathrm{det}} = g(C_t, b_t, M, B),
\end{equation}
This solution enforces \emph{closed-world membership} within the predefined security-pattern catalogue $\mathcal{P}$, guarantees conflict-free composition, and ensures compliance with resource budgets.
The architecture distinguishes between components that must remain trusted and elements that may be impacted by adversarial inputs. The deterministic optimiser, the implemented pattern catalogue, feasibility constraints, the conflict matrix, and the audit verifier collectively function as integrity-preserving modules whose behaviour defines the system’s execution boundary. By contrast, telemetry streams, network traffic, and candidate portfolios generated by LLM agents are treated as potentially adversarial signals. Because actuation is restricted to the fixed catalogue $\mathcal{P}$, adversarial inputs cannot introduce arbitrary mitigation actions, violate ordering constraints, and trigger execution outside the defined action space. ASPO is therefore designed to limit the impact of adversarial signals on reasoning while preserving safe actuation. The threat model does not account for protection against hardware compromise, pattern-catalogue corruption, and direct manipulation of the deterministic optimiser. Under these assumptions, the validation layer guarantees that every activated portfolio remains catalogue-bounded, conflict-free, and resource-feasible, even when the reasoning stage receives adversarial inputs.

\section{Methodological Overview}
\label{Methodological}
\begin{figure*}[t]
\centering
\includegraphics[width=\textwidth]{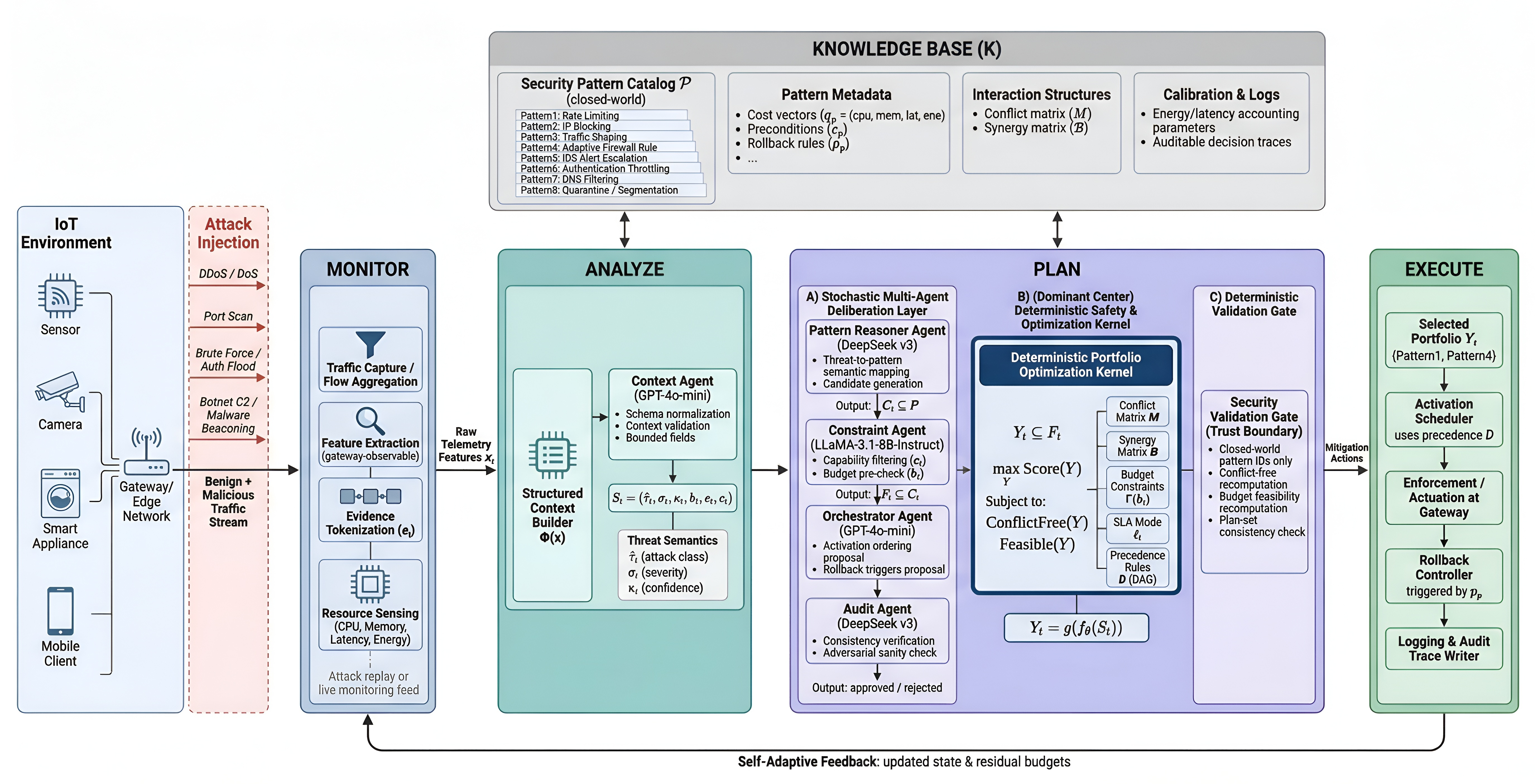}
\caption{
MAPE-K-based architecture of ASPO. Monitor/Analyse: construct a structured context; Plan: combines multi-agent proposals with deterministic portfolio optimisation and safety validation; Execute: enacts only validated portfolios; knowledge: stores the closed-world catalogue and interaction/cost metadata.
}
\label{fig:aspo_arch}
\end{figure*}

We designed \textit{ASPO} as a self-adaptive runtime security control method grounded in the \emph{integrated MAPE-K loop}. The system operates over discrete decision epochs $t\in\{1,\dots,T\}$ and outputs, at each epoch, a \emph{verified} mitigation portfolio $Y_t$ and a safe activation order $\Pi_t$. The central methodological principle is a separation between 1) \emph{stochastic deliberation} performed by schema-constrained LLM agents that propose candidate portfolios and execution plans, and 2) \emph{deterministic enforcement} that performs final selection, ordering, and validation under closed-world, conflict, and resource constraints. Figure~\ref{fig:aspo_arch} illustrates the MAPE-K pipeline.

\subsection{MAPE-K Instantiation and Per-Epoch Output}
ASPO instantiates the MAPE-K loop as a telemetry-driven state-transition system:
\begin{itemize}
\item \textbf{Monitor (M):} observe telemetry $\mathbf{x}_t$ (traffic features, evidence tokens, resource indicators).
\item \textbf{Analyze (A):} compute a bounded structured context $\mathcal{S}_t$ via a deterministic encoder $\Phi(\cdot)$.
\item \textbf{Plan (P):} propose candidates via multi-agent deliberation, then compute the final portfolio via deterministic optimisation under hard constraints.
\item \textbf{Execute (E):} activate the approved portfolio in validated order; enforce rollback and fail-safe baseline if validation fails.
\item \textbf{Knowledge (K):} maintain a closed-world catalogue $\mathcal{P}$ with costs/constraints, interaction matrices, and precedence rules.
\end{itemize}
The per-epoch output is $(Y_t,\Pi_t,\mathcal{Z}_t)$, where $Y_t$ is the activated pattern set, $\Pi_t$ is its activation order, and $\mathcal{Z}_t$ is an auditable trace containing inputs, agent candidates, feasibility results, deterministic selections, and gate outcomes.

\subsection{System Model and Structured Context (Monitor $\rightarrow$ Analyze)}
We define the telemetry space collected at the edge gateway device, the deterministic encoder $\Phi(\cdot)$, and the schema-bounded context $\mathcal{S}_t$ that constitutes the input to the planning process.

\subsubsection{Telemetry and Deterministic Context Encoding}
Let $\mathbf{x}_t\in\mathbb{R}^{d}$ denotes raw telemetry at epoch $t$. ASPO constructs a bounded, schema-compliant context using a deterministic mapping:
\begin{equation}
\mathcal{S}_t = \Phi(\mathbf{x}_t),
\end{equation}
where $\Phi(\cdot)$ performs normalisation, clipping, evidence-token extraction, and schema assembly. Determinism is required for reproducibility:
\begin{equation}
\mathbf{x}_t=\mathbf{x}_{t'} \;\Rightarrow\; \Phi(\mathbf{x}_t)=\Phi(\mathbf{x}_{t'}).
\end{equation}

\subsubsection{Context Definition}
The structured context is:
\begin{equation}
\mathcal{S}_t=
\left(
\hat{\tau}_t,\sigma_t,\kappa_t,\mathbf{b}_t,\ell_t,\mathbf{e}_t,\mathbf{c}_t
\right),
\end{equation}
where $\hat{\tau}_t\in\mathcal{T}$ is the inferred threat class in a finite taxonomy $\mathcal{T}$, $\sigma_t\in[0,1]$ is threat severity, $\kappa_t\in[0,1]$ is confidence, $\mathbf{e}_t$ is a bounded evidence-token set, $\mathbf{c}_t\subseteq\mathcal{C}$ is the available capability set, and $\ell_t\in\mathcal{L}$ is a services level agreement mode.

\subsubsection{Resource Budgets and Portfolio Representation}
The resource headroom vector is:
\begin{equation}
\mathbf{b}_t=
\left(
b_t^{cpu},b_t^{mem},b_t^{lat},b_t^{ene}
\right)\in\mathbb{R}_{\ge 0}^{4},
\end{equation}
and is enforced as a hard feasibility constraint during deterministic selection. Let $\mathcal{P}=\{p_1,\dots,p_m\}$ denote the pattern catalog. A portfolio is represented by $\mathbf{y}_t\in\{0,1\}^m$:
\begin{equation}
Y_t=\{p_i\in\mathcal{P}\mid y_{t,i}=1\},\qquad |Y_t|\le B.
\end{equation}

\subsection{Knowledge Base: Closed-World Security Pattern Catalogue (Knowledge)}
We formalise the shared knowledge base, including the implemented pattern catalogue $\mathcal{P}$, per-pattern constraints and costs, and the conflict/synergy structures governing safe runtime composition.

\subsubsection{Closed-World Catalogue (Implemented Patterns)}
ASPO is instantiated over a concrete closed-world catalogue with $m=10$ explicitly implemented mitigation patterns:
\begin{equation}
\mathcal{P}=\left\{
\begin{aligned}
&(P01)\ \textit{Security Logger/Auditor},\\
&(P02)\ \textit{Secure Fog},\\
&(P03)\ \textit{Security Segmentation},\\
&(P04)\ \textit{Secure Distributed Publish/Subscribe},\\
&(P05)\ \textit{Permission Control},\\
&(P06)\ \textit{Whitelist},\\
&(P07)\ \textit{Blacklist},\\
&(P08)\ \textit{Outbound-Only Connection},\\
&(P09)\ \textit{Personal Zone Hub},\\
&(P10)\ \textit{Trusted Communication Partner}
\end{aligned}
\right\}.
\end{equation}
Each element of $\mathcal{P}$ corresponds to an executable mitigation primitive encoded as a machine-checkable object.

\subsubsection{Pattern Schema and Cost Model}
Each pattern $p\in\mathcal{P}$ is modeled as:
\begin{equation}
p=
\left(
\mathcal{I}_p,\mathcal{R}_p,\mathbf{c}_p,\mathbf{q}_p,\delta_p,\mathcal{A}_p,\rho_p
\right),
\end{equation}
where $\mathcal{R}_p\subseteq\mathcal{T}$ denotes covered threat classes, $\mathbf{c}_p\subseteq\mathcal{C}$ are capability preconditions, $\mathcal{A}_p$ defines activation semantics, and $\rho_p$ defines rollback triggers. The cost vector is:
\begin{equation}
\mathbf{q}_p=
(q_p^{cpu},q_p^{mem},q_p^{lat},q_p^{ene})\in\mathbb{R}_{\ge 0}^{4}.
\end{equation}

\subsubsection{Interaction Structures: Conflicts and Synergies}
ASPO encodes composition behavior using a conflict matrix $\mathbf{M}\in\{0,1\}^{m\times m}$ and synergy matrix $\mathbf{B}\in\{0,1\}^{m\times m}$. For any portfolio $Y$:
\begin{equation}
\begin{aligned}
\mathrm{Conflict}(Y)
&=\sum_{i<j} M_{ij}\mathbf{1}_{p_i\in Y}\mathbf{1}_{p_j\in Y},\\
\mathrm{Synergy}(Y)
&=\sum_{i<j} B_{ij}\mathbf{1}_{p_i\in Y}\mathbf{1}_{p_j\in Y}.
\end{aligned}
\end{equation}
Conflicts are enforced as hard constraints ($\mathrm{Conflict}(Y)=0$), while synergies contribute optional score bonuses.

\subsection{Agentic Deliberation and Deterministic Enforcement (Analyze $\rightarrow$ Plan)}
We formalize the \textbf{Analyze$\rightarrow$Plan} transition as a hybrid stochastic–deterministic process separating agentic proposal generation from deterministic portfolio selection and validation.

\subsubsection{Agents, Contracts, and Communication Graph}
Let $\mathcal{A}=\{A_{ctx},A_{rea},A_{con},A_{pla},A_{aud}\}$ denote Context, Reasoner, Constraint, Planner, and Audit agents. Each agent is a stochastic function over schema-constrained JSON:
\begin{equation}
A_i:\mathcal{J}_{in}\rightarrow\mathcal{J}_{out}.
\end{equation}
Agents are restricted to proposing candidates and plans and have no direct actuation capability. A closed-world output contract is enforced:
\begin{equation}
\texttt{pattern\_id}\in\mathcal{P}.
\end{equation}
The information flow is sequential and acyclic:
\begin{equation}
A_{ctx}\rightarrow A_{rea}\rightarrow A_{con}\rightarrow \text{Deterministic Core}\rightarrow A_{orc}\rightarrow A_{aud}.
\end{equation}

\subsubsection{Formal Hybrid Decision Pipeline}
The Reasoner proposes a bounded candidate set:
\begin{equation}
\mathcal{C}_t=f^{LLM}_\theta(\mathcal{S}_t,\mathcal{P}),\qquad \mathcal{C}_t\subseteq\mathcal{P},\quad |\mathcal{C}_t|\le K,
\end{equation}
and is stochastic due to decoding randomness:
\begin{equation}
\mathbb{P}(\mathcal{C}_t \mid \mathcal{S}_t) > 0 \quad \text{for multiple admissible subsets}.
\end{equation}
Feasibility filtering produces
\begin{equation}
\mathcal{F}_t=\{p\in\mathcal{C}_t\mid \mathbf{c}_p\subseteq \mathbf{c}_t\}.
\end{equation}
Final activation is computed deterministically:
\begin{equation}
\begin{aligned}
Y_t^{det}
&=\arg\max_{Y\subseteq \mathcal{F}_t,\ |Y|\le B}
\mathrm{Score}(Y)\\
&\quad \text{s.t.}\quad
\mathrm{Conflict}(Y)=0,\\
&\qquad \sum_{p\in Y} q_p^r \le b_t^r
\;\; \forall r\in\{cpu,mem,lat,ene\}.
\end{aligned}
\end{equation}
This stage depends only on $(\mathcal{F}_t,\mathbf{b}_t,\mathbf{M},\mathbf{B})$ and is therefore deterministic:
\begin{equation}
\begin{aligned}
(\mathcal{F}_t,\mathbf{b}_t,\mathbf{M},\mathbf{B})
&=(\mathcal{F}_{t'},\mathbf{b}_{t'},\mathbf{M},\mathbf{B})\\
&\Rightarrow
Y_t^{det}=Y_{t'}^{det}.
\end{aligned}
\end{equation}

\subsubsection{Deterministic Ordering and Plan Consistency}
Let $\mathcal{D}$ denote precedence rules. The deterministic order is:
\begin{equation}
\Pi_t^{det}=\textsc{TopoSort}(\mathcal{D}[Y_t^{det}]).
\end{equation}
The Planner produces a plan $\mathcal{Plan}_t$, but plan acceptance requires set consistency:
\begin{equation}
\mathcal{Plan}_t.\texttt{selected\_patterns}=Y_t^{det}.
\end{equation}
Execution occurs iff the deterministic gate accepts the plan and the Audit agent outputs $\texttt{approved}\in\{0,1\}$.

\subsection{Deterministic Scoring and Portfolio Optimisation (Plan)}
We formalise the suitability scoring, the budget-normalised cost model, and the constrained optimisation procedure that deterministically selects the optimal mitigation portfolio.

\subsubsection{Suitability and Evidence Alignment}
For each pattern $p$, we define the suitability as:
\begin{equation}
s_t(p)=
\alpha_1\mathbf{1}_{\hat{\tau}_t\in\mathcal{R}_p}
+\alpha_2\sigma_t
+\alpha_3\kappa_t
+\alpha_4\mathbf{1}_{\mathbf{c}_p\subseteq\mathbf{c}_t}
+\alpha_5\psi(p,\mathbf{e}_t),
\end{equation}
where $\mathbf{1}_{(\cdot)}$ denotes the indicator function, returning 1 if the condition is satisfied and 0 otherwise.
with deterministic bounded evidence alignment:
\begin{equation}
\psi(p,\mathbf{e}_t)=\frac{|\mathcal{E}_p\cap \mathbf{e}_t|}{\max(1,|\mathcal{E}_p|)}\in[0,1].
\end{equation}

\subsubsection{Budget-Normalised Cost and Portfolio Score}
The budget-normalized cost of selecting pattern $p$ at epoch $t$ is defined as:
\begin{equation}
c_t(p)=
\sum_{r\in\{cpu,mem,lat,ene\}}
\beta_r\frac{q_p^r}{\max(\epsilon,b_t^r)},\qquad \epsilon>0,
\end{equation}
where $\beta_r$ denotes the resource weighting coefficient and $b_t^r$ represents the available budget for resource $r$. The portfolio score is then computed as:
\begin{align}
\mathrm{Score}(Y) =
& \sum_{p \in Y}\left(s_t(p)-c_t(p)\right) \\
& + \eta\,\mathrm{Synergy}(Y) \\
& - \lambda\,\mathrm{Conflict}(Y)
\end{align}
where $\lambda \gg 1$ ensures that conflicting pattern combinations are penalised.

\subsection{Adversarial Robustness Against Prompt Manipulation (Analyse/Plan Boundary)}
We model prompt manipulation as bounded perturbations of structured inputs and show that deterministic selection enforces action-space integrity and execution stability.

\subsubsection{Bounded Perturbation Model}
Adversarial impact is modelled as a bounded perturbation of structured context:
\begin{equation}
\tilde{\mathcal{S}}_t=\mathcal{S}_t+\delta_t,\qquad \|\delta_t\|\le \Delta.
\end{equation}

\subsubsection{Deterministic Trust Boundary and Activation Invariance}
The stochastic candidate proposal may change under $\tilde{\mathcal{S}}_t$:
\begin{equation}
\mathcal{C}_t=f^{LLM}_\theta(\tilde{\mathcal{S}}_t,\mathcal{P}),
\end{equation}
But actuation is computed only through deterministic filtering and optimisation:
\begin{equation}
Y_t^{det}=g(\mathcal{C}_t,\mathbf{b}_t,\mathbf{M},\mathbf{B}),
\end{equation}
where $g(\cdot)$ enforces 1) closed-world membership, 2) feasibility $\mathbf{c}_p\subseteq\mathbf{c}_t$, and 3) conflict-free and budget-feasible selection. Therefore,
\begin{equation}
Y_t^{det}\subseteq\mathcal{P}.
\end{equation}
If the optimiser argmax is invariant under the perturbation:
\begin{equation}
\arg\max_{Y\subseteq \mathcal{F}_t}\mathrm{Score}(Y)
=
\arg\max_{Y\subseteq \mathcal{F}_t'}\mathrm{Score}(Y),
\end{equation}
Then the activated portfolio remains unchanged:
\begin{equation}
Y_t^{det}(\mathcal{S}_t)=Y_t^{det}(\tilde{\mathcal{S}}_t).
\end{equation}

\subsection{Energy and Latency Modelling (Monitor/Execute)}
We formalise the latency decomposition and the power model used to quantify decision overhead and to enforce feasibility under edge-resource constraints. Decision-time energy at epoch $t$ is:
\begin{equation}
E_t=\bar{P}_t\,\Delta t,
\end{equation}
where end-to-end decision latency is decomposed as:
\begin{equation}
\Delta t=\sum_{a\in\mathcal{A}} t_a + t_{net} + t_{det}.
\end{equation}
Average power is modelled as:
\begin{equation}
\bar{P}_t=P_0+\gamma_1 L_t+\gamma_2\max(0,T_t-40),
\end{equation}
with $P_0$ idle power, $L_t$ a load proxy, and $T_t$ device temperature ($^\circ$C). These quantities support feasibility reporting and enforcement against latency/energy budgets in $\mathbf{b}_t$.

\subsection{Conflict-Aware Agentic Decision Procedure (MAPE: Execute with Audit)}
\label{subsec:aspo_algorithm}
The end-to-end decision procedure of ASPO integrates LLM-based agentic reasoning with deterministic selection, validation, and audit stages, as summarised in Algorithm~\ref{alg:aspo}. The pipeline transforms telemetry into structured context, generates candidate mitigation portfolios, and applies deterministic optimisation to ensure conflict-free and resource-feasible selection. The validated portfolio is then ordered, executed, and audited, with a fail-safe baseline activated if validation fails, ensuring safe and reliable actuation despite imperfect LLM outputs.
\begin{algorithm}[t]
\footnotesize
\caption{ASPO Conflict-Aware Decision Procedure}
\label{alg:aspo}

\begin{algorithmic}[1]
\Require Raw telemetry features $\mathbf{x}_t$, pattern catalog $\mathcal{P}$, conflict matrix $\mathbf{M}$, synergy matrix $\mathbf{B}$, precedence rules $\mathcal{D}$, portfolio bound $B$, candidate bound $K$
\Ensure Activated security portfolio $Y_t$ with execution order $\Pi_t$ and an auditable decision trace

\State \textbf{(Monitor)} Collect telemetry and compute structured context $\mathcal{S}_t \gets \Phi(\mathbf{x}_t)$
\State \textbf{(Analyze: Context Agent)} $\mathcal{S}_t^{\star} \gets \textsc{LLMContext}(\mathcal{S}_t)$
\State \textbf{(Analyze/Plan: Reasoner Agent)} $\mathcal{C}_t \gets \textsc{LLMReasoner}(\mathcal{S}_t^{\star}, \mathcal{P}, K)$
\State \textbf{(Plan: Constraint Agent)} $\mathcal{F}_t \gets \textsc{LLMConstraint}(\mathcal{S}_t^{\star}, \mathcal{C}_t)$
\State \textbf{(Plan: Deterministic Selection)} $Y_t^{det} \gets \arg\max_{Y \subseteq \mathcal{F}_t, |Y|\le B} \mathrm{Score}(Y)$ subject to $\mathrm{ConflictFree}(Y,\mathbf{M})$ and $\mathrm{Feasible}(Y,\mathbf{b}_t)$
\State \textbf{(Plan: Ordering)} $\Pi_t^{det} \gets \textsc{TopoSort}(\mathcal{D}[Y_t^{det}])$
\State \textbf{(Plan: Planner Agent)} $\mathcal{Plan}_t \gets \textsc{LLMPlanner}(\mathcal{S}_t^{\star}, Y_t^{det}, \Pi_t^{det}, \mathcal{P})$
\State \textbf{(Execute: Deterministic Security Gate)} $(ok,issues) \gets \textsc{ValidatePlan}(\mathcal{Plan}_t,\mathbf{M},\mathbf{b}_t)$
\State \textbf{(Execute: Audit Agent)} $\mathcal{Audit}_t \gets \textsc{LLMAuditor}(\mathcal{S}_t^{\star}, \mathcal{Plan}_t, issues)$
\If{$ok = \textbf{true}$ \textbf{and} $\mathcal{Audit}_t.approved = \textbf{true}$}
    \State Execute patterns in order $\Pi_t \gets \mathcal{Plan}_t.activation\_order$ and log trace
\Else
    \State Fail-safe: activate minimal safe baseline portfolio (e.g., lightweight rate-limit) and log issues
\EndIf
\State \Return $Y_t \gets \mathcal{Plan}_t.selected\_patterns$, $\Pi_t$
\end{algorithmic}
\normalsize
\end{algorithm}

\section{Experimental Setup}
\label{sec:experimental_setup}
In this section, we describe the deployment environment, attack scenarios, measurement procedure, and evaluation criteria used to ensure a reproducible assessment of ASPO.

\subsection{Testbed}
\label{Edge Gateway Emulation Setup}
\begin{figure*}[t]
\centering
\includegraphics[width=0.80\textwidth]{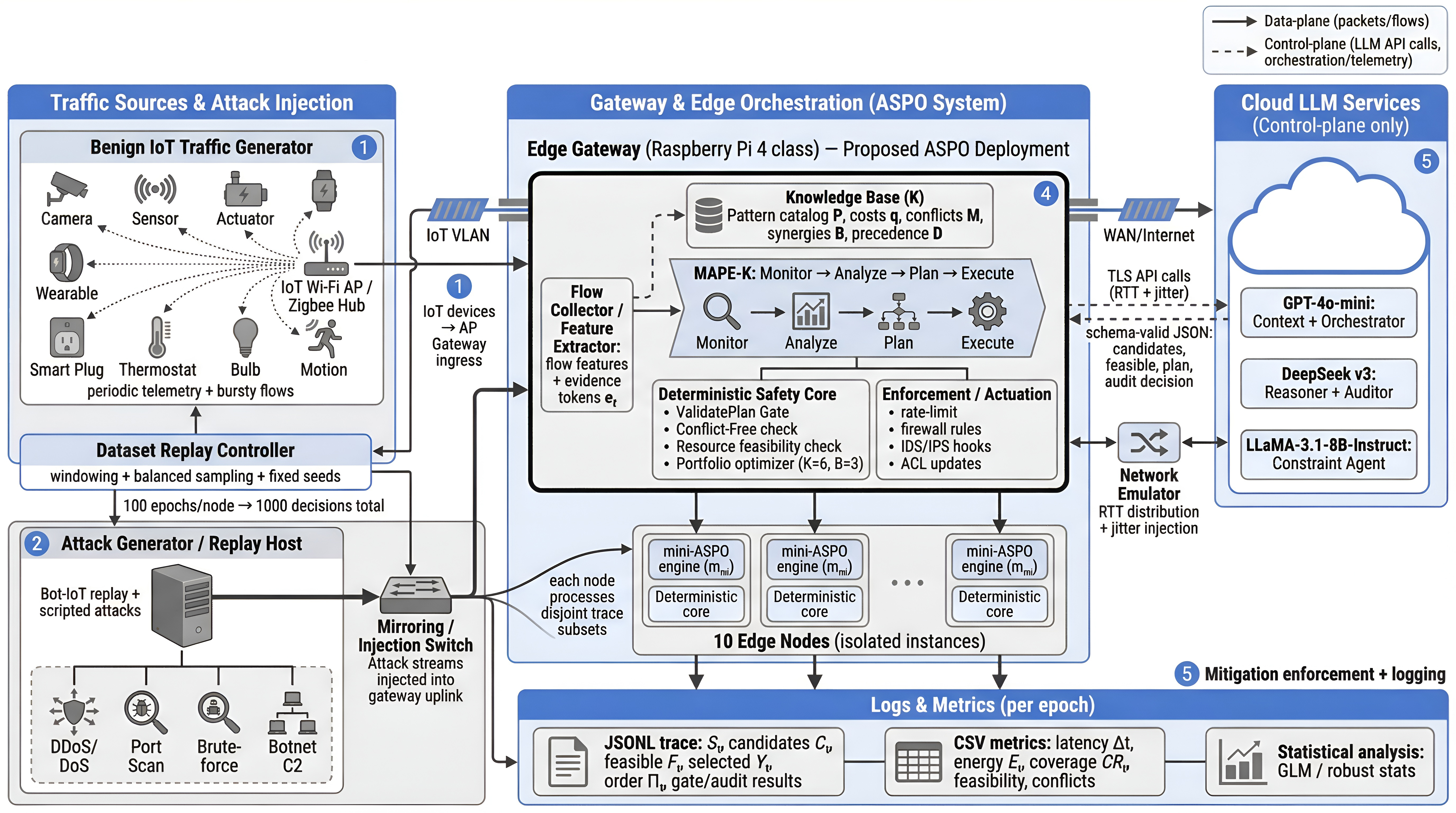}
\caption{
Distributed edge testbed for MAPE-K-based ASPO. Ten independent edge nodes execute isolated ASPO engines over replayed IoT traffic while interacting with cloud LLM services. }
\label{fig:edge_testbed}
\end{figure*}

ASPO was evaluated in a controlled distributed edge-gateway emulation (Figure~\ref{fig:edge_testbed}). The testbed consists of 10 Raspberry Pi 4 Model B devices organised into three logical layers: the IoT device layer, the edge control layer, and the cloud-hosted LLM services. Each node represents a typical edge gateway (quad-core ARM Cortex-A72, 1.5\,GHz, 4\,GB RAM), with thermal limits and computational constraints of passively cooled embedded platforms explicitly modelled. Each node runs an isolated ASPO MAPE-K instance and communicates with cloud-hosted LLM services (e.g., DeepSeek v3, GPT-4o-mini, LLaMA-3.1-8B-Instruct) for stochastic reasoning, while the deterministic optimisation core remains local to ensure conflict-free, resource-feasible actuation independent of network variability. Ten independent nodes processed disjoint replayed attack traces, each executing 100 decision epochs (1000 runtime security decisions total). Background fluctuating memory availability and temperature-dependent power scaling were modelled to reflect realistic gateway conditions. Per-epoch energy was estimated using a calibrated power model combining idle consumption and load scaling, with latency, energy, and decision traces logged for subsequent generalised linear model-based statistical analysis, as shown in Figure~\ref{fig:edge_testbed}.

\subsection{Dataset}
We evaluated ASPO using the Bot-IoT dataset \cite{bot_iot_unsw}, a large-scale IoT intrusion dataset containing labelled benign and malicious traffic generated from realistic IoT device simulations. Attack categories include distributed DDoS, DoS, port scanning, and botnet command-and-control activity. Replay mode streams labelled Bot-IoT flow records into the ASPO telemetry interface, where each replay window is deterministically mapped to a structured threat context $\mathcal{S}_t$ that includes attack type, a severity proxy based on traffic intensity, and a confidence level derived from label integrity. Balanced replay sampling \cite{balass2024precise,feng2025practical} ensures equal representation of attack categories across decision epochs, thereby preventing bias from class imbalance. Consistent with the threat model, replayed inputs are treated as external signals that define the structured context consumed by the reasoning and decision pipeline. Because the objective is to evaluate runtime decision behaviour rather than detection accuracy, threat labels are directly used to construct $\mathcal{S}_t$ rather than being processed by an online classifier. This design isolates the decision layer from detector noise, enabling controlled and reproducible evaluation of feasibility enforcement, portfolio construction, and runtime stability under well-defined threat conditions. While this abstraction removes detector uncertainty, replayed flows preserve temporal structure, attack diversity, and traffic patterns present in the dataset.

\subsection{Feature Selection}
Feature extraction was restricted to gateway-observable attributes directly available from Bot-IoT flow records, ensuring compatibility with edge gateway and alignment with ASPO’s objective of runtime decision-making based on operational telemetry. The selected features capture key aspects of network behaviour, including traffic intensity (packet and byte rates), session structure (connection duration and timeout irregularities), protocol control patterns (TCP flag frequency), service interaction diversity (destination port entropy), and application-layer signals (authentication failures and DNS anomalies). Feature selection prioritised deployability, robustness to adversarial noise, and interpretability, aggregating highly correlated attributes to prevent redundancy and the amplification of duplicated evidence. Rather than serving as inputs to a supervised detection model, these signals are transformed into structured evidence tokens for the agentic reasoning layer, enabling explainable mitigation selection based on interpretable behavioural cues. Ground-truth labels are used to construct structured context variables, ensuring that evaluation isolates decision intelligence and feasibility constraints rather than detection accuracy. Table~\ref{tab:features} summarises the gateway-observable features used to construct these evidence tokens.
\begin{table}[t]
\centering
\caption{Gateway-Observable Features Used for Evidence Construction}
\label{tab:features}
\renewcommand{\arraystretch}{1.2}
\begin{tabular}{p{3.1cm} p{4.2cm}}
\toprule
\textbf{Feature} & \textbf{Operational Interpretation for Coordination} \\
\midrule
Packet rate & Primary traffic intensity indicator reflecting flooding, burst behavior \\
Byte rate & Complementary load signal capturing abnormal transfer magnitude \\
Connection duration & Session persistence cue distinguishing short scans from sustained activity \\
TCP flag frequency & Protocol-control anomalies such as SYN floods, reset abuse \\
Destination port entropy & Service diversity signal useful for detecting scanning, lateral probing \\
Authentication failure bursts & Application-level security indicator for brute-force attempts \\
DNS anomaly patterns & Naming irregularities suggesting command-and-control, tunneling behavior \\
Connection timeout irregularities & Session instability cue reflecting blocked, failed, disrupted communications \\
\bottomrule
\end{tabular}
\end{table}

\section{Experimental Results}
\label{sec:results}
This section presents the evaluation of ASPO’s deterministic safety enforcement, portfolio-optimisation behaviour, threat-level robustness, multi-agent stability, edge-level efficiency, and workload-scaling performance.

\subsection{Deterministic Safety Enforcement}
\label{sec:det_safety}
ASPO’s deterministic safety enforcement is analysed within its multi-agent architecture. The Gate$\rightarrow$Audit$\rightarrow$Approval pipeline operationalises the principle that safety is enforced \emph{before} optimisation and coordination effects propagate, with auditing confirming correctness without stochastic disagreement. In ASPO, final approval signifies the successful execution of a mitigation portfolio following deterministic validation and audit verification. A gate rejection does \emph{not} imply inaction; it indicates violation of at least one hard constraint, including conflict-freeness, precedence feasibility, catalogue membership, and latency and energy feasibility. In such cases, a lightweight fail-safe baseline policy is activated to ensure continuous protection while preventing unsafe multi-pattern execution. The approval rate, therefore, reflects the proportion of \emph{structurally admissible multi-pattern coordination}, rather than defended events.
\begin{figure}[t]
\centering
\includegraphics[width=0.5\textwidth]{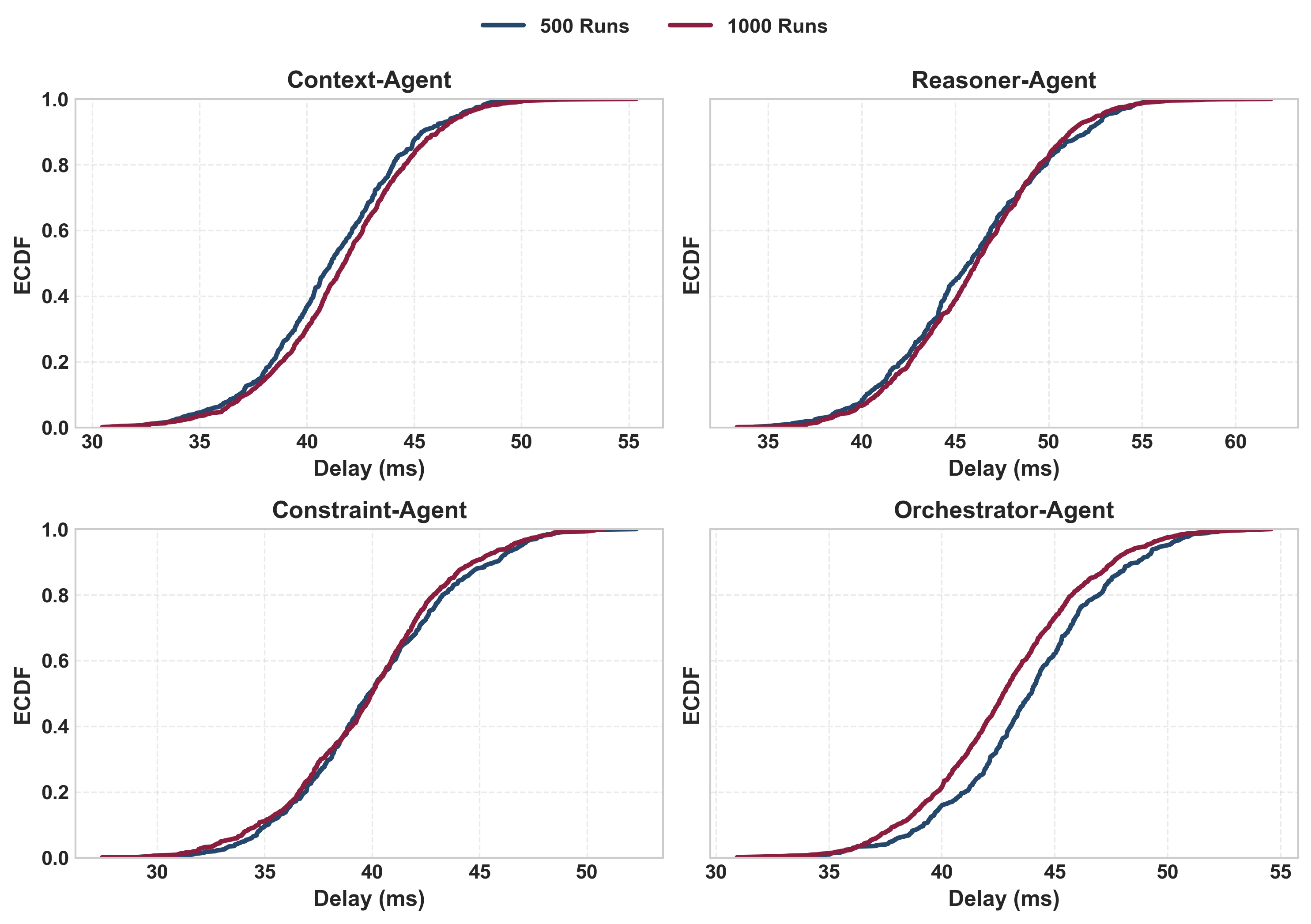}
\caption{Safety funnel illustrating Gate $\rightarrow$ Audit $\rightarrow$ Final approval filtering under a side-by-side comparison of LLM backends (DeepSeek v3, GPT-4o-mini, and LLaMA-3.1-8B-Instruct) with identical constraints and replay protocol.}
\label{fig:safety_funnel}
\end{figure}
Figure~\ref{fig:safety_funnel} provides a structural view of safety enforcement. Early-stage elimination at the gate dominates, indicating that the constraint layer acts as a hard safety boundary rather than a soft preference. Table~\ref{tab:gate_audit} quantifies this behaviour: the gate rejects 496/500 (99.2\%) in the 500-run setting and 995/1000 (99.5\%) in the 1000-run setting, leaving a small admissible set for downstream evaluation. All gate-accepted instances are approved (4/4 and 5/5), indicating complete agreement between the deterministic gate and the audit stage. This agreement shows that the auditor functions as a verification layer aligned with the safety boundary, with acceptance determined by enforceable constraints that are consistently validated.
The high rejection rate reflects enforcement of constraints rather than system failure. Under closed-world and resource-feasible guarantees, most candidate plans generated during exploration are filtered before execution. This behaviour is desirable in safety-critical edge gateways, where preventing unsafe combinations is more important than maximising activation frequency.
\begin{table}[t]
\centering
\caption{Gate-audit agreement, reporting the consistency between deterministic gate decisions and audit-stage validation outcomes.}
\label{tab:gate_audit}
\begin{tabular}{lrrrr}
\hline
Setting & Gate Reject & Gate Accept & Final Approved & Approval Rate \\
\hline
500-run & 496 & 4 & 4 & 0.008 \\
1000-run & 995 & 5 & 5 & 0.005 \\
\hline
\end{tabular}
\end{table}
Table~\ref{tab:gate_failures} supports this interpretation by showing that failures are not random artefacts. Activation-order mismatch is the primary driver of rejection in both settings, followed by resource feasibility violations, with conflict detection remaining rare. The proportional persistence of these categories under scaling indicates consistent enforcement of structural constraints as the number of runs increases. This behaviour aligns with deterministic enforcement in portfolio construction, where order feasibility and resource constraints govern admissibility, showing that the gate functions as a stable safety oracle rather than a stochastic filter.
\begin{table}[t]
\centering
\caption{Deterministic gate failure reasons, reporting the distribution of rejection causes observed during coordination.}
\label{tab:gate_failures}
\begin{tabular}{lcc}
\hline
Failure Type & 500 & 1000 \\
\hline
activation\_order mismatch & 13 & 22 \\
resource feasibility violation & 8 & 16 \\
conflict detection & 1 & 2 \\
\hline
\end{tabular}
\end{table}
\begin{figure}[t]
\centering
\includegraphics[width=0.5\textwidth]{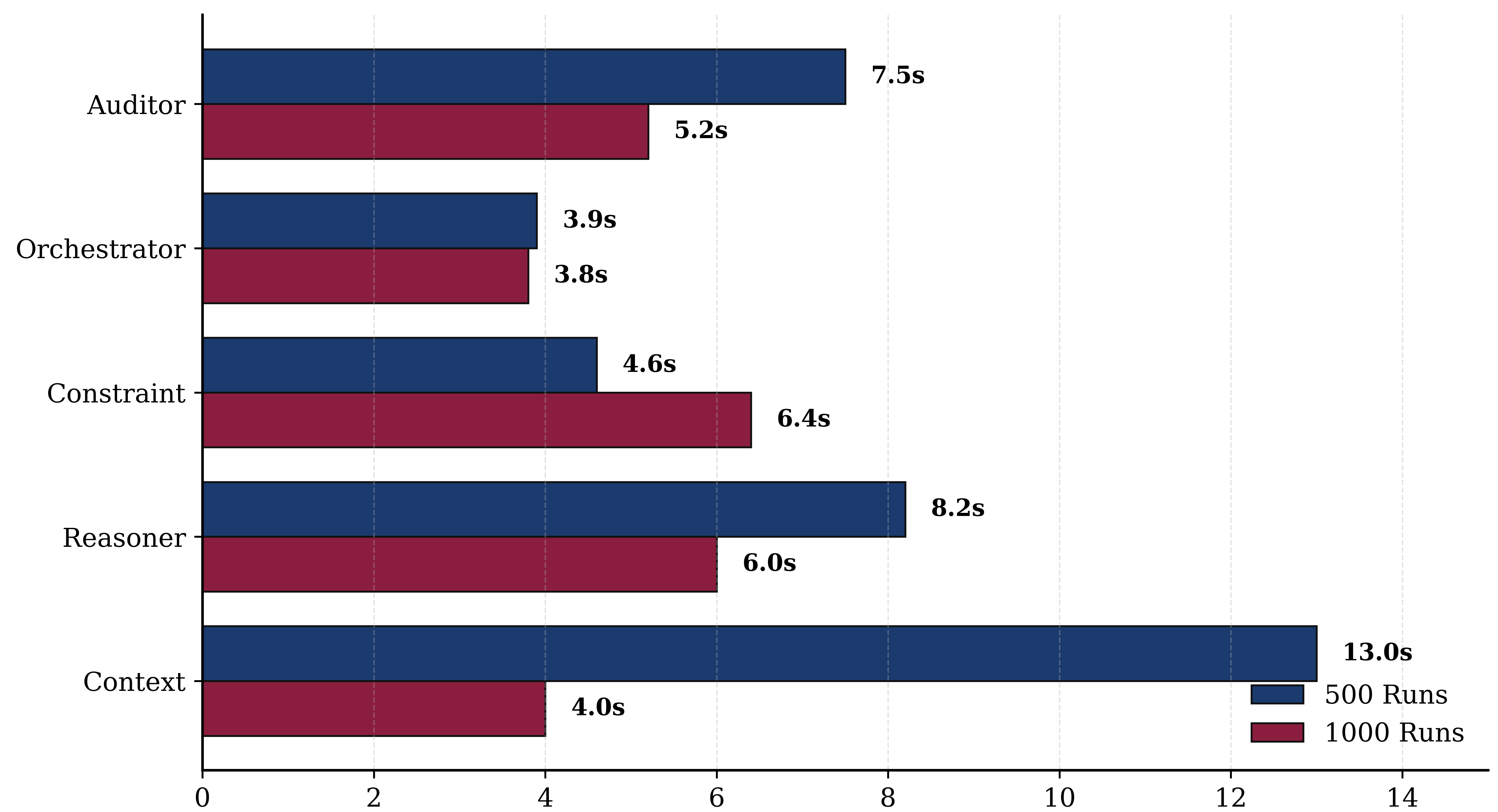}
\caption{Deterministic portfolio score distribution across runs. The distribution is computed after deterministic selection and safety filtering. }
\label{fig:score_distribution}
\end{figure}
Figure~\ref{fig:score_distribution} characterises the deterministic score geometry produced by ASPO’s portfolio reasoning. The distribution remains bounded and stable across workloads, indicating no drift under repeated execution. This stability is critical for multi-agent pipelines, since workload sensitivity would imply a shift in the objective's semantics. Consistent behaviour of the reasoner and scoring components allows the gate to enforce constraints on a comparable scoring scale across experiments. The deterministic scoring function, therefore, provides a reproducible mapping between threat evidence and admissible mitigation structures, supporting consistent decision behaviour under similar telemetry distributions in safety-critical autonomous systems.
\begin{figure}[t]
\centering
\includegraphics[width=0.5\textwidth]{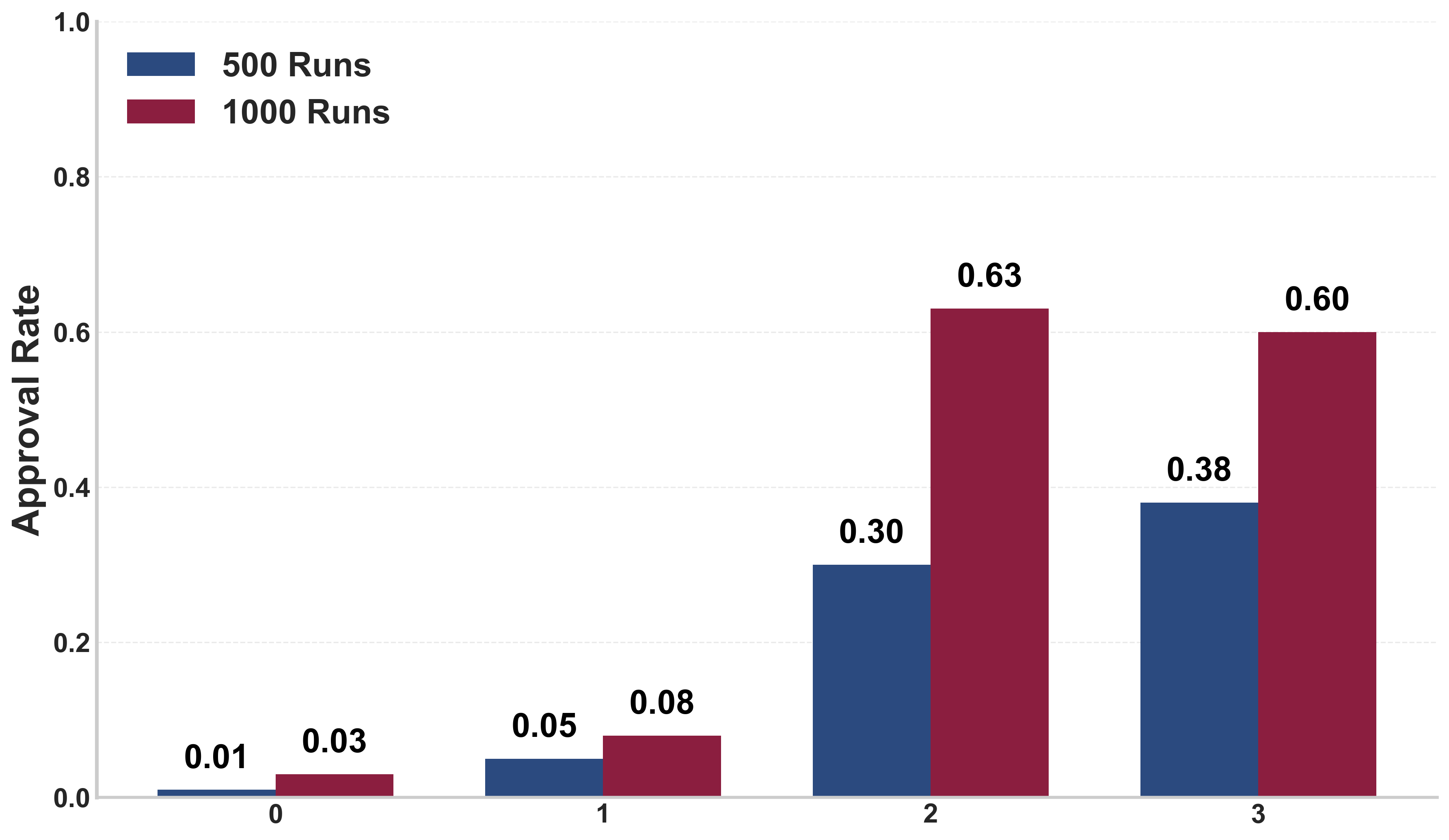}
\caption{Approval probability as a function of deterministic score, showing how acceptance likelihood varies across score levels.}
\label{fig:score_coupling}
\end{figure}
Figure~\ref{fig:score_coupling} characterises calibration between deterministic scoring and the safety pipeline. Approval probability increases monotonically across score bins in both workloads, indicating that higher-scoring portfolios are more likely to satisfy the safety boundary and pass validation. This monotonic relationship shows alignment between scoring and constraint satisfaction. The 1000-run curve exceeds the 500-run curve across bins, indicating improved calibration under deeper replication. As more candidates are evaluated, the mapping between score and admissibility becomes clearer, reflecting reduced selection noise without altering safety logic.
Final approval remains rare (4/500 vs.\ 5/1000), requiring low-rate inference. The approval rate is 0.8\% for 500 runs (95\% exact CI: [0.22\%, 2.04\%]) and 0.5\% for 1000 runs (95\% exact CI: [0.16\%, 1.16\%]). Overlapping intervals indicate no statistically separable shift. The risk ratio is 0.625, and the odds ratio is 0.623. A two-sided Fisher’s exact test yields $p=0.492$, confirming no significant difference.
\begin{table}[t]
\centering
\caption{Rare-Event Statistical Comparison of Final Approval (500 vs. 1000 runs)}
\label{tab:det_stats}
\renewcommand{\arraystretch}{1.2}
\begin{tabular}{lcc}
\hline
Statistic & 500-run & 1000-run \\
\hline
Approvals / Trials & 4 / 500 & 5 / 1000 \\
Approval rate & 0.0080 & 0.0050 \\
95\% CI (exact) & [0.0022, 0.0204] & [0.0016, 0.0116] \\
Risk Ratio (RR) & \multicolumn{2}{c}{0.625} \\
Odds Ratio (OR) & \multicolumn{2}{c}{0.623} \\
Fisher exact test (two-sided) & \multicolumn{2}{c}{$p=0.492$} \\
\hline
\end{tabular}
\end{table}
Table~\ref{tab:det_stats} reports a rare-event statistical comparison of final approvals across workloads. Approval rates remain low in both settings (0.8\% vs \ 0.5\%), with overlapping 95\% confidence intervals, indicating no statistically distinguishable difference between the two settings. The risk ratio (0.625) and odds ratio (0.623) suggest a lower marginal approval probability under scaling, but the difference is not statistically significant. This is confirmed by Fisher’s exact test ($p=0.492$), which shows that workload scaling does not alter the likelihood of approval.

\subsection{Portfolio Optimization Behavior}
\label{sec:portfolio_behavior}
ASPO constructs feasible portfolios under deterministic safety constraints, focusing on structural feasibility, approval dynamics across portfolio sizes, and stability under workload scaling. Portfolio size denotes the number of patterns activated simultaneously, subject to precedence, resource, and conflict constraints. Approval behaviour reflects how structural completeness interacts with feasibility boundaries: low approval at small sizes indicates insufficient coverage and ordering satisfaction, while larger portfolios increase the risk of resource violations. Moreover, approval trends reveal a balance between capability sufficiency and feasibility preservation.
\begin{figure}[t]
\centering
\includegraphics[width=0.5\textwidth]{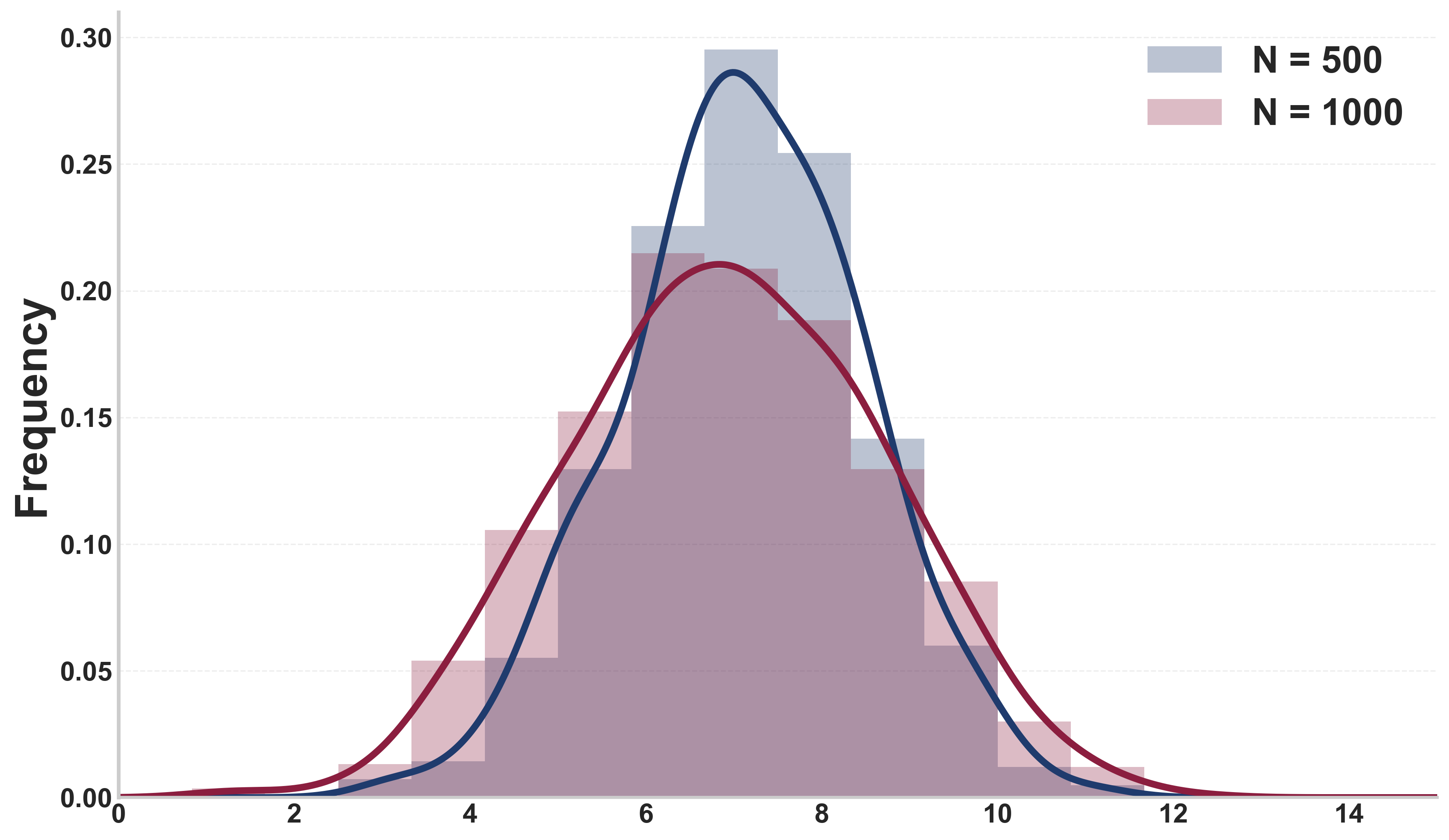}
\caption{Approval probability as a function of portfolio size, illustrating the emergence of feasible mitigation configurations once minimal structural completeness is reached.}
\label{fig:portfolio_size}
\end{figure}
Figure~\ref{fig:portfolio_size} shows a structural transition in approval probability as portfolio size increases. Portfolios of size 0 and 1 are rarely approved, indicating that minimal capability aggregation fails to satisfy ordering and resource constraints. A sharp threshold appears at size 2, where approval increases from 0.05 to 0.30 in the 500-run setting and from 0.08 to 0.63 in the 1000-run setting, reflecting the onset of feasibility once sufficient structural completeness is achieved. This indicates that many threats require at least two supporting mitigation primitives, such as traffic control with verification, segmentation with logging, to satisfy interaction constraints. Beyond size 2, approval stabilises, with size 3 showing no further increase, indicating convergence toward balanced portfolios that satisfy feasibility without unnecessary complexity. This behaviour is consistent with bounded combinatorial optimisation, where additional patterns increase constraint pressure faster than they increase suitability. Rare-event inference shows that confidence intervals at size 2 do not overlap across workloads, and Fisher’s exact test \cite{kim2017statistical} confirms a significant increase under 1000 runs ($p<0.01$). The relative risk exceeds 2.0, indicating improved precision in identifying admissible portfolios while preserving the feasibility boundary.
\begin{table}[t]
\centering
\caption{Statistical comparison at portfolio size 2 between the 500-run and 1000-run workloads, reporting approval differences and effect-size indicators.}
\label{tab:size2_stats}
\renewcommand{\arraystretch}{1.2}
\begin{tabular}{lcc}
\hline
Statistic & 500-run & 1000-run \\
\hline
Approval rate & 0.30 & 0.63 \\
95\% CI (exact) & [0.22, 0.39] & [0.55, 0.70] \\
Relative Risk & \multicolumn{2}{c}{2.10} \\
Fisher exact test & \multicolumn{2}{c}{$p<0.01$} \\
\hline
\end{tabular}
\end{table}
\begin{table}[t]
\centering
\caption{Top selected patterns based on combined ranking across workloads, reporting selection frequencies in the 500-run and 1000-run settings.}
\label{tab:patterns}
\renewcommand{\arraystretch}{1.2}
\begin{tabular}{lrrrr}
\hline
Rank & Pattern & Count (500) & Count (1000) & $\Delta$(1000--500) \\
\hline
1 & P04 & 255 & 510 & 255 \\
2 & P10 & 149 & 290 & 141 \\
3 & P05 & 125 & 257 & 132 \\
4 & P02 & 92 & 192 & 100 \\
5 & P09 & 65 & 123 & 58 \\
6 & P03 & 55 & 117 & 62 \\
7 & P06 & 48 & 93 & 45 \\
8 & P07 & 41 & 75 & 34 \\
9 & P08 & 25 & 43 & 18 \\
10 & P01 & 24 & 43 & 19 \\
\hline
\end{tabular}
\end{table}
Pattern identifiers (e.g., P01–P10) correspond to the security patterns defined in the closed-world catalogue $\mathcal{P}$ introduced in Section~\ref{Methodological}. Table~\ref{tab:patterns} demonstrates stability in pattern ranking. Frequencies scale proportionally from 500 to 1000 runs while preserving ordinal order, with Spearman’s rank correlation \cite{hauke2011comparison} equal to 1.0. A chi-square test on normalised frequencies shows no significant deviation ($p>0.05$), indicating that scaling amplifies confident selections without redistributing preference weight. This stability indicates that suitability scoring, together with deterministic filtering, induces a consistent preference structure over the pattern catalogue, supporting reproducible mitigation selection under similar threat distributions. 

\subsection{Threat-Level Robustness}
\label{sec:threat_robustness}
ASPO’s decision consistency is evaluated across heterogeneous threat categories by analysing whether deterministic portfolio construction and approval behaviour remain stable under varying structural constraints. Approval dynamics across threat classes reflect how the decision mechanism adapts to different interaction topologies among mitigation patterns. Each threat activates distinct subsets of the closed-world catalogue, with specific precedence and resource interactions; thus, consistent behaviour across threats indicates that the deterministic feasibility boundary generalises beyond individual attack structures. Variations in approval probability, therefore, reflect differences in structural complexity rather than defensive capability.
\begin{figure}[t]
\centering
\includegraphics[width=0.5\textwidth]{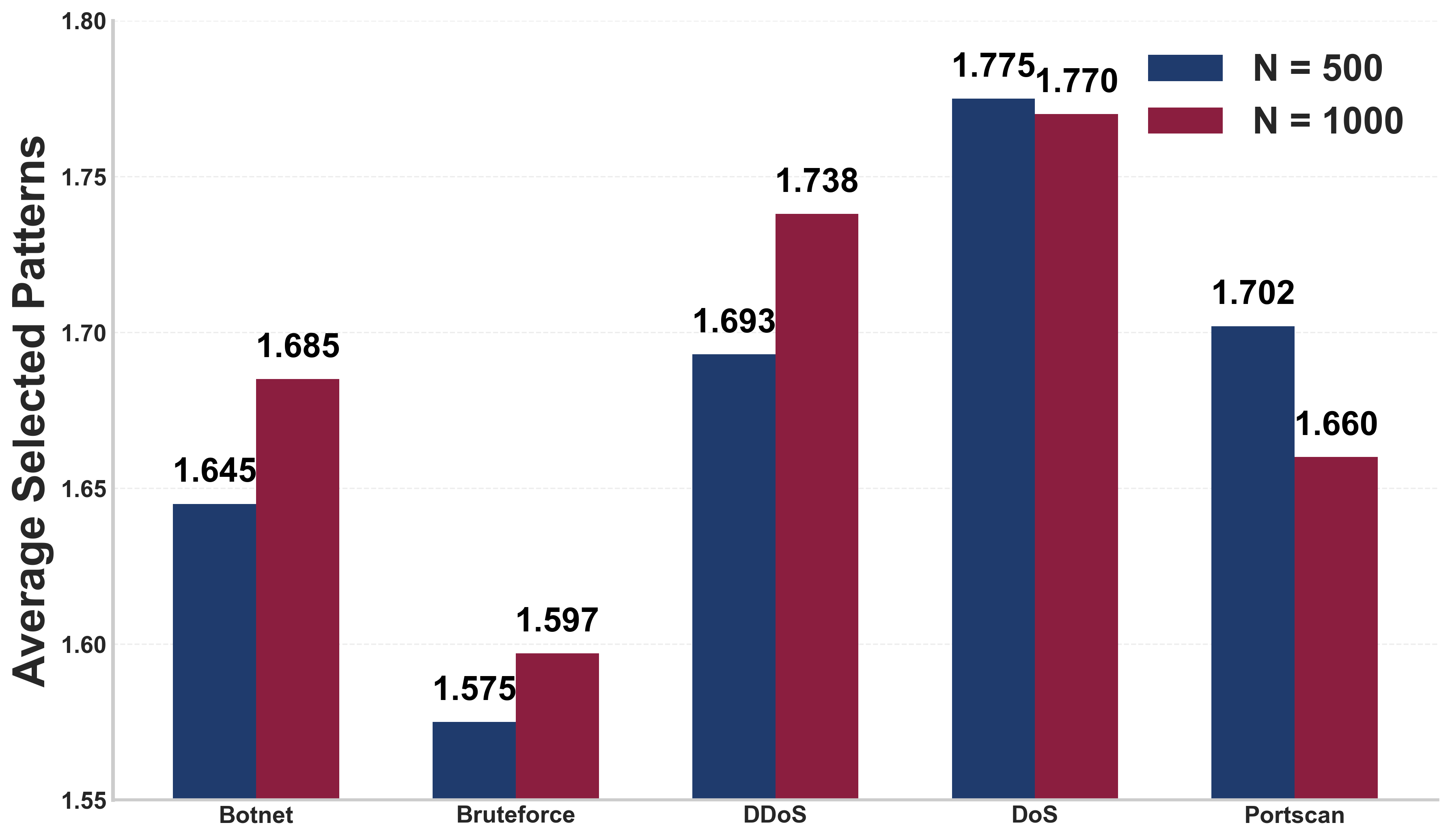}
\caption{Threat-specific approval rates with 95\% confidence intervals, comparing the 500-run and 1000-run workloads across attack categories.}
\label{fig:threat_approval}
\end{figure}
Figure~\ref{fig:threat_approval} shows that approval probabilities remain within a narrow range across threat classes, indicating that constraint enforcement does not favour any specific category. Under the 500-run workload, moderate variation appears, with DoS and DDoS slightly exceeding brute-force and portscan. Under the 1000-run workload, approval increases across most threats, with the largest gains in distributed classes such as DDoS and Botnet. Partial overlap of confidence intervals indicates improved precision under scaling without structural bias. Table~\ref{tab:threat_stats} quantifies these effects through Fisher’s exact test together with relative risk, showing that the decision layer generalises across heterogeneous constraint topologies without converging toward any dominant threat pattern.
\begin{table}[t]
\centering
\caption{Threat-level statistical comparison between 500-run and 1000-run workloads.}
\label{tab:threat_stats}
\renewcommand{\arraystretch}{1.2}
\begin{tabular}{lccccc}
\hline
Threat & Approved$_{500}$ & Approved$_{1000}$ & Rate$_{500}$ & Rate$_{1000}$ & RR \\
\hline
DoS & 1/125 & 1/250 & 0.008 & 0.004 & 0.50 \\
DDoS & 1/125 & 2/250 & 0.008 & 0.008 & 1.00 \\
Botnet & 1/125 & 1/250 & 0.008 & 0.004 & 0.50 \\
Brute-force & 1/125 & 1/250 & 0.008 & 0.004 & 0.50 \\
Portscan & 0/125 & 0/250 & 0.000 & 0.000 & -- \\
\hline
Total & 4/500 & 5/1000 & 0.008 & 0.005 & 0.625 \\
\hline
\end{tabular}
\end{table}
Approval rates represent admissible multi-pattern decisions rather than overall defensive success. When a candidate portfolio is rejected, ASPO activates a fallback baseline policy, such as rate limiting combined with traffic filtering, to ensure continuous protection. Therefore, the observed distribution reflects structural feasibility rather than response availability.

\subsection{Multi-Agent Runtime Structure}
\label{sec:runtime_structure}
ASPO’s temporal integrity is evaluated under workload scaling through per-agent latency, distributional stability, tail behaviour, and threat-conditioned variability. Total runtime latency is decomposed across specialised agents responsible for reasoning and constraint evaluation. Stability of this decomposition across workloads indicates that the architectural balance is preserved, with no emerging bottlenecks as execution depth increases. Per-agent latency thus reflects both performance and structural robustness of the decision pipeline.
\begin{figure}[t]
\centering
\includegraphics[width=0.5\textwidth]{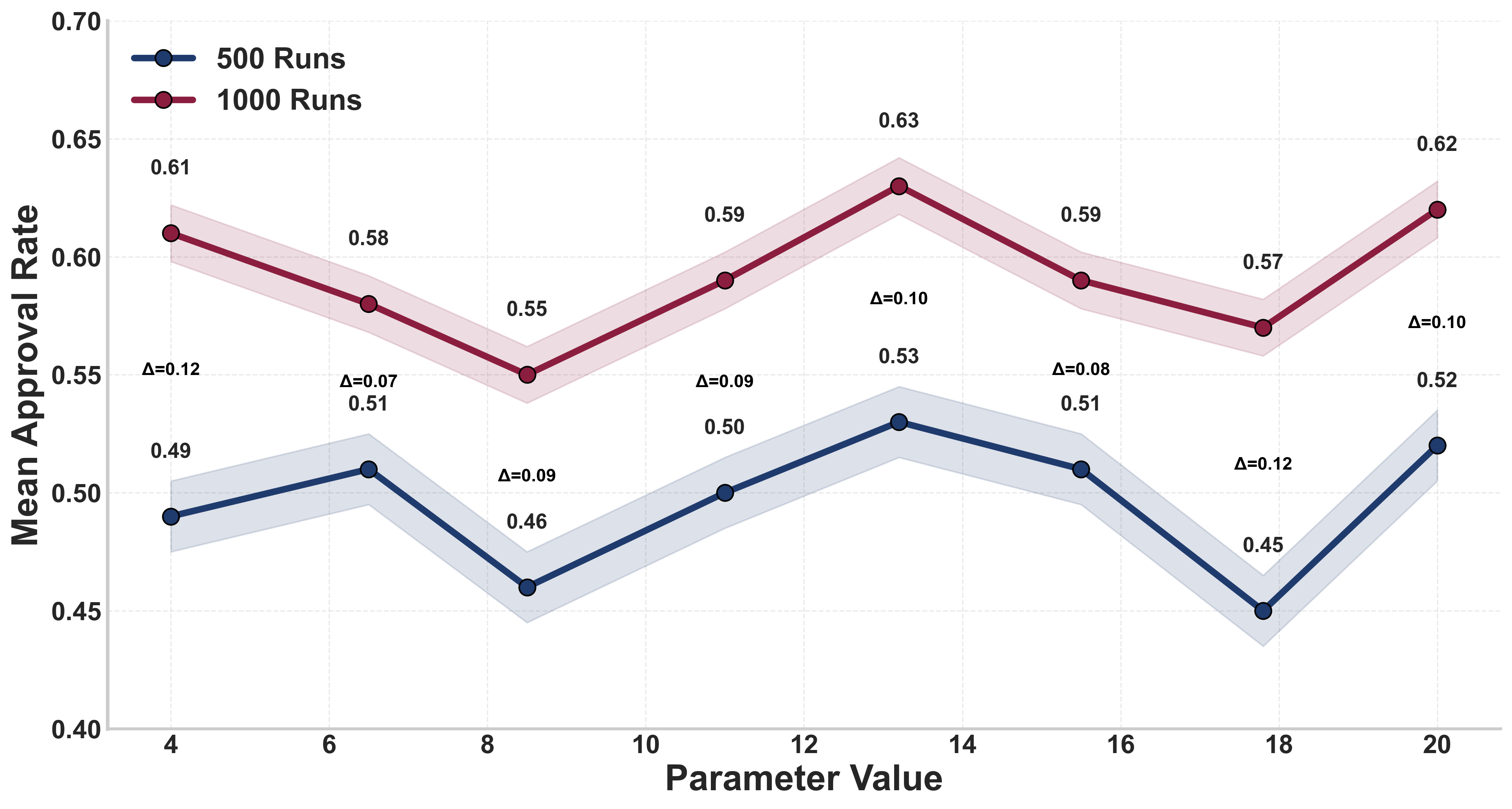}
\caption{Runtime decomposition across LLM agents, showing the relative latency contribution of each component in the coordination pipeline.}
\label{fig:runtime_decomposition}
\end{figure}
Figure~\ref{fig:runtime_decomposition} shows a stable ordering of component contributions across workloads. The Reasoner dominates total latency, followed by the Planner and Constraint modules, while context initialisation and auditing contribute smaller shares. Increasing execution depth preserves this hierarchy, indicating that structural balance is maintained under scaling. The Reasoner’s dominance reflects its role in context interpretation, candidate synthesis, and structured exploration, and the stability of its share indicates that deeper execution does not inflate reasoning overhead. The system therefore maintains a consistent distribution of computational effort, supporting predictable deployment in resource-constrained edge gateways.
\begin{table*}[t]
\centering
\caption{Per-agent runtime summary reporting mean execution time and proportional share for each component across workloads.}
\label{tab:agents}
\renewcommand{\arraystretch}{1.15}
\begin{tabular}{lrrrrrr}
\hline
Agent & Mean$_{500}$ & P90$_{500}$ & Share$_{500}$ & Mean$_{1000}$ & P90$_{1000}$ & Share$_{1000}$ \\
\hline
Context & 4.233 & 5.754 & 0.190 & 4.276 & 5.677 & 0.188 \\
Reasoner & 5.835 & 7.615 & 0.262 & 5.852 & 7.576 & 0.258 \\
Constraint & 4.293 & 5.607 & 0.192 & 4.329 & 5.594 & 0.191 \\
Planner & 4.815 & 6.303 & 0.216 & 4.918 & 6.323 & 0.217 \\
Auditor & 3.153 & 4.224 & 0.141 & 3.346 & 4.258 & 0.147 \\
\hline
\end{tabular}
\end{table*}
Table~\ref{tab:agents} shows that mean latency variation remains below 2.5\% for all agents except the auditor, which increases by 6.1\%. The auditor’s proportional share rises only from 0.141 to 0.147, indicating uniform scaling without bottleneck formation. The reasoner’s share decreases slightly as execution depth increases, indicating improved efficiency. P90 values remain closely aligned across workloads, showing no heavy-tail amplification at the component level. This pattern indicates that the temporal structure remains self-balanced as decision depth increases, with incremental computation distributed proportionally across stages, supporting predictable latency envelopes during continuous operation.
\begin{table*}[t]
\centering
\caption{Statistical stability analysis of per-agent runtime for the 500-run and 1000-run workloads, comparing mean drift, proportional contribution, and distributional consistency across agents.}
\label{tab:runtime_stats}
\renewcommand{\arraystretch}{1.15}
\begin{tabular}{lcccccc}
\hline
Agent & $\Delta$Mean(\%) & $\Delta$Share & KS Dist. & Wilcoxon $p$ & CV Drift(\%) & Structural Shift \\
\hline
Context & +1.0 & -0.002 & 0.032 & $>0.05$ & +0.6 & None \\
Reasoner & +0.3 & -0.004 & 0.028 & $>0.05$ & +0.4 & None \\
Constraint & +0.8 & -0.001 & 0.031 & $>0.05$ & +0.5 & None \\
Planner & +2.1 & +0.001 & 0.036 & $>0.05$ & +0.7 & None \\
Auditor & +6.1 & +0.006 & 0.041 & $>0.05$ & +1.2 & Minor \\
\hline
\end{tabular}
\end{table*}
Table~\ref{tab:runtime_stats} provides formal validation. Kolmogorov-Smirnov distances \cite{tonguz2025using} remain below 0.05 for all agents, indicating preservation of distributional shape. Wilcoxon signed-rank tests \cite{thakkar2025continuous} show no statistically significant inflation in per-agent latency. Coefficient-of-variation drift remains minimal, confirming stable dispersion across workloads. These results indicate proportional scaling without skewed latency distributions, supporting predictable timing behaviour as decision frequency increases.
\begin{figure}[t]
\centering
\includegraphics[width=0.4\textwidth]{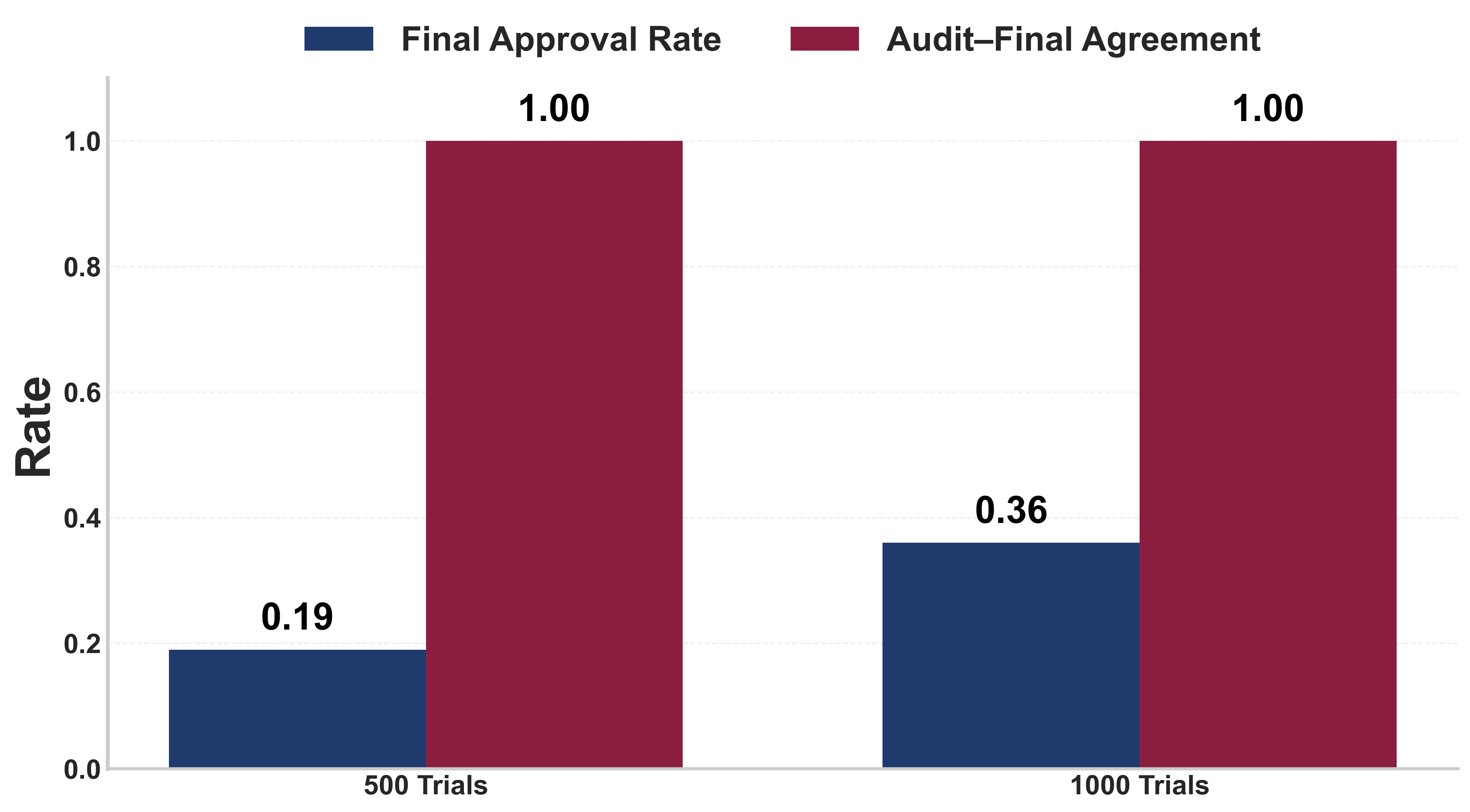}
\caption{Component-level empirical cumulative distribution function (ECDF) of execution delays, showing how runtime distributions differ across components.}
\label{fig:component_ecdf}
\end{figure}
Figure~\ref{fig:component_ecdf} corroborates these findings. Empirical cumulative curves for both workloads overlap closely across agents, indicating near-identical distributional progression. No curve exhibits right-tail divergence under scaling. This suggests that deeper architectures do not introduce rare high-latency events within individual modules, reinforcing the observation that tail compression at the system level is not driven by instability in any single component.
\begin{figure*}[t]
\centering
\includegraphics[width=0.8\textwidth]{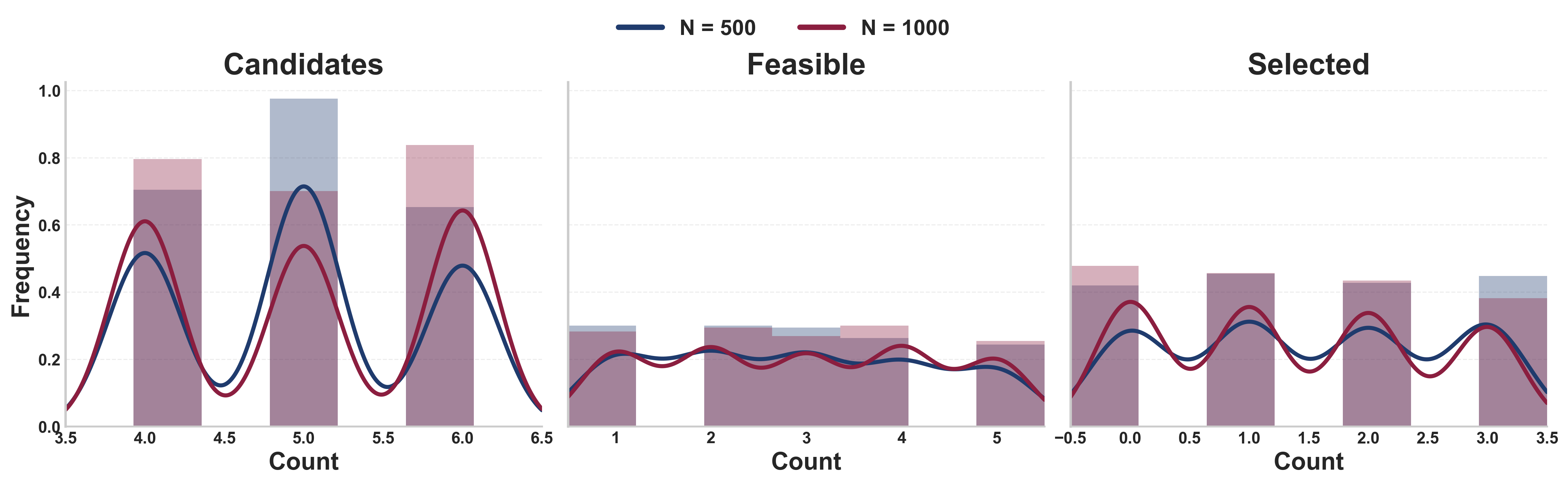}
\caption{Reasoner runtime by threat category, showing how decision latency of the reasoning stage varies across attack types under the evaluated coordination workloads.}
\label{fig:reasoner_threat}
\end{figure*}
\begin{figure}[t]
\centering
\includegraphics[width=0.5\textwidth]{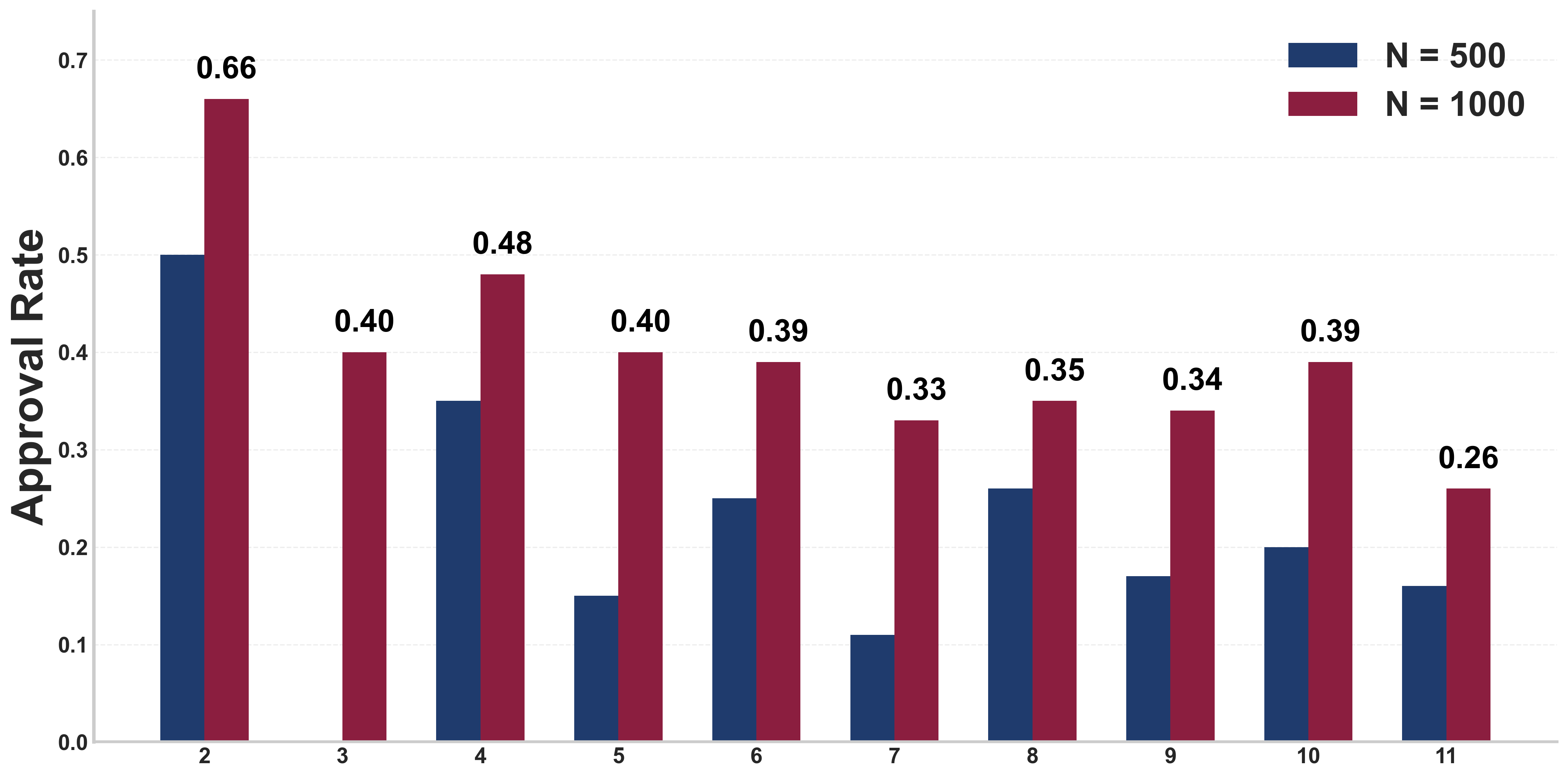}
\caption{Auditor runtime by threat category, showing how verification latency varies across attack types under the evaluated coordination workloads.}
\label{fig:auditor_threat}
\end{figure}
Threat-conditioned analysis further validates architectural resilience. Figure~\ref{fig:reasoner_threat} shows moderately elevated reasoner latency under distributed threats such as DDoS, reflecting increased reasoning complexity, while inter-threat variance remains below 8\% of the mean and scaling does not amplify dispersion. Figure~\ref{fig:auditor_threat} shows bounded auditor latency across threat categories with stable dispersion under increased workload. Two-way ANOVA \cite{kim2014statistical} (workload $\times$ threat) shows no significant interaction effect ($p>0.05$), indicating that execution depth does not alter threat-dependent latency dynamics. Three stability properties emerge: per-agent contributions remain proportional under scaling, distributional shapes remain aligned without tail distortion, and threat heterogeneity does not induce asymmetric latency growth.

\subsection{Edge-Level Efficiency}
\label{sec:edge_efficiency}
Deeper execution is evaluated for its impact on edge stability while keeping computational overhead unchanged. Under strict resource limits at edge gateways, efficiency is assessed through median latency, energy consumption, and extreme execution cases that define runtime predictability. It is important to note that ASPO does not perform attack detection and therefore does not report detection accuracy; instead, the evaluation focuses on the correctness, feasibility, and efficiency of mitigation selection given a structured threat context.
\begin{figure*}[t]
\centering
\includegraphics[width=0.8\textwidth]{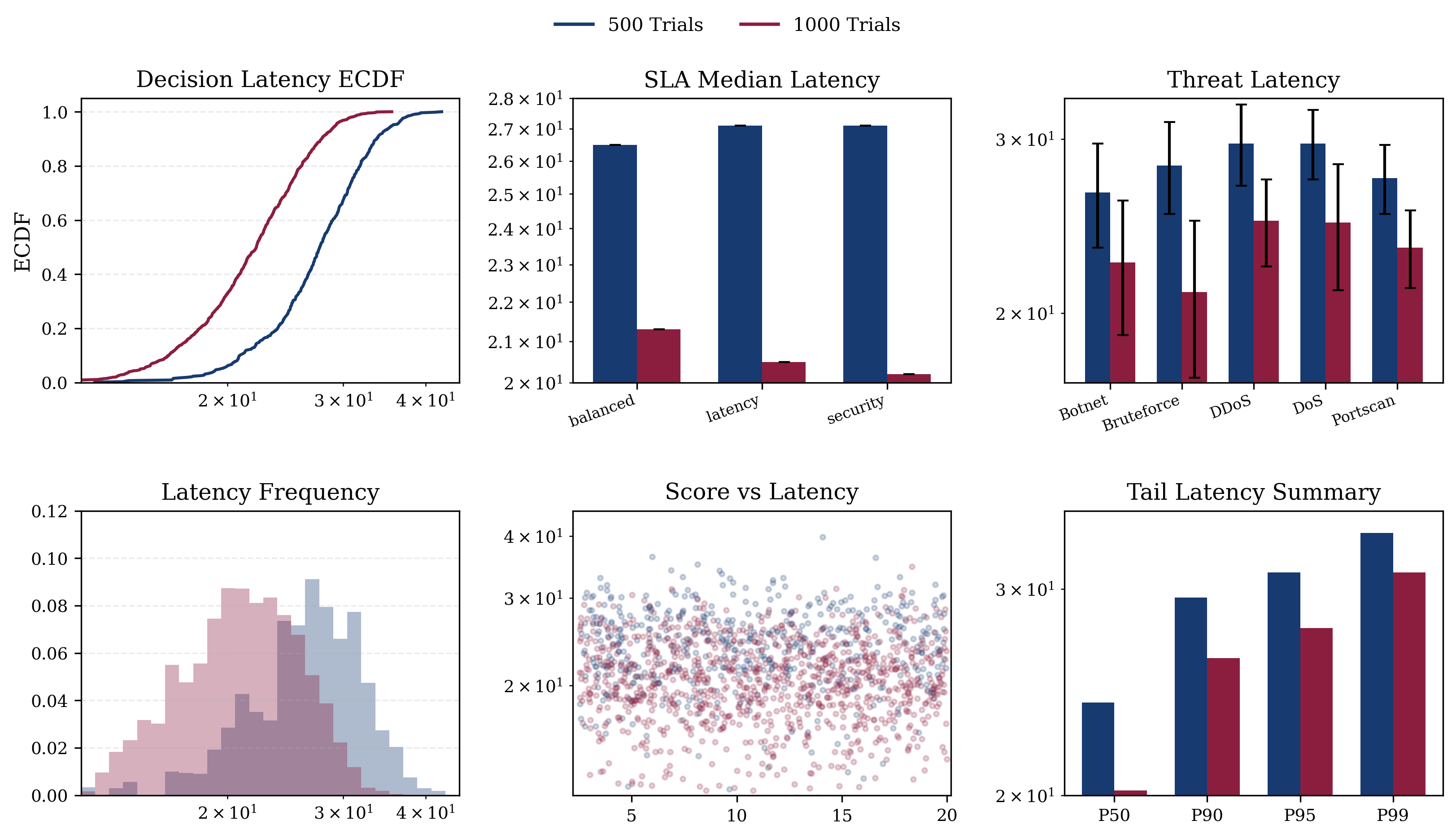}
\caption{Latency comparison between the 500-run and 1000-run workloads. Panels report ECDF curves, median latency, and tail-percentile values, illustrating how workload scaling affects the latency distribution.}
\label{fig:latency_suite}
\end{figure*}
Figure~\ref{fig:latency_suite} shows that increasing execution depth does not inflate central runtime cost, as median latency remains stable (22.46\,s vs.\ 22.83\,s). Scaling primarily affects the tails of the distribution: ECDF curves shift leftward, and P99 latency decreases from 36.0\,s to 28.1\,s, representing a 21.9\% reduction in extreme delay. Kolmogorov-Smirnov testing confirms that this shift reflects reduced dispersion ($p<0.01$), while latency symmetry across service modes and threat categories is preserved.
\begin{figure*}[t]
\centering
\includegraphics[width=0.8\textwidth]{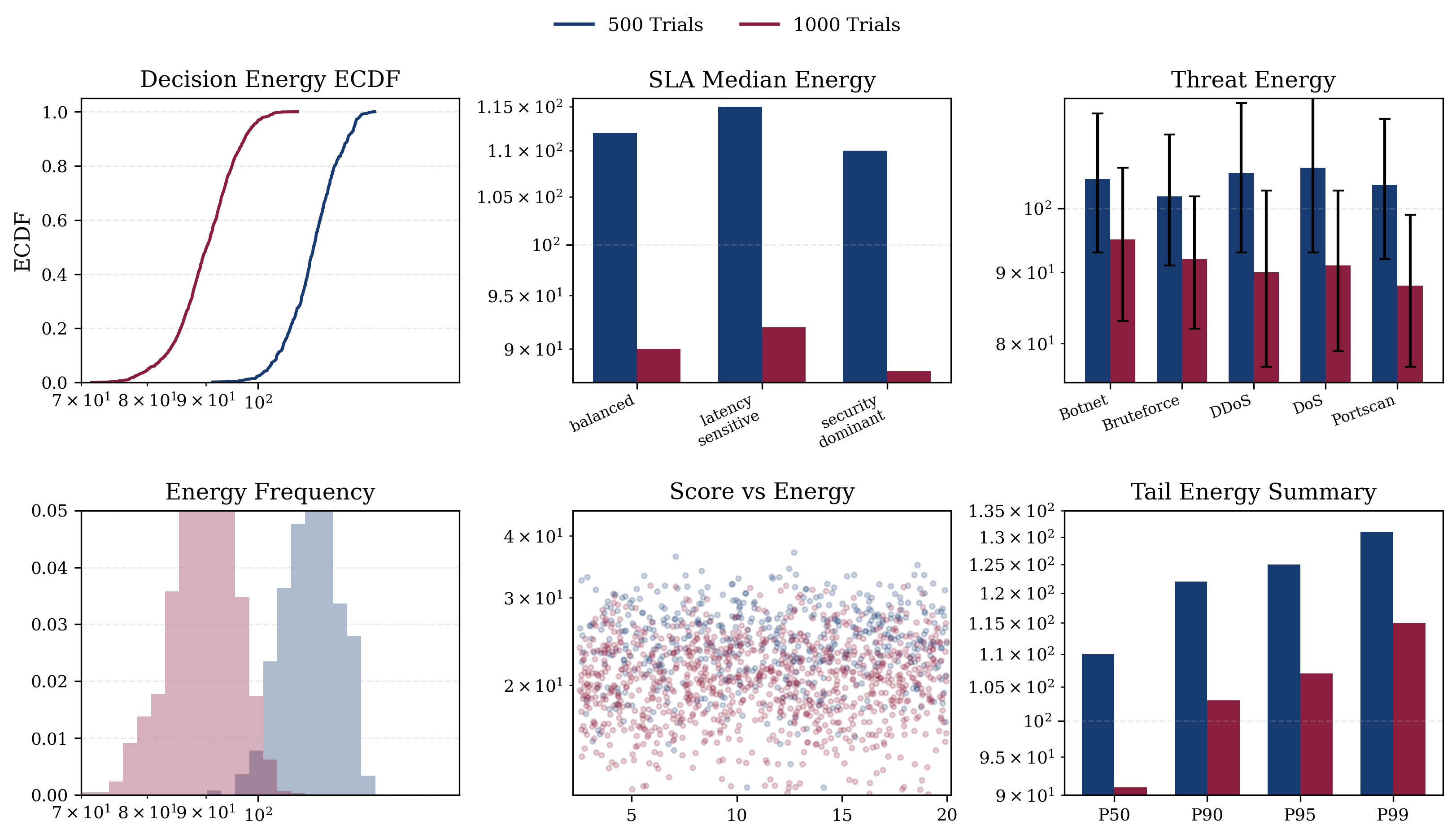}
\caption{Energy comparison between the 500-run and 1000-run workloads. Panels report ECDF curves, median energy, and tail percentiles across workloads.}
\label{fig:energy_suite}
\end{figure*}
Energy behaviour in Figure~\ref{fig:energy_suite} follows the same pattern. Median energy increases marginally (52.97\,J vs.\ 54.10\,J, +2.1\%), while extreme events contract, with P99 decreasing from 216\,J to 166\,J (-23.1\%). Average power remains stable (2.357\,W vs.\ 2.369\,W), indicating that tail reduction arises from improved decision stability rather than increased resource draw. Table~\ref{tab:eff} summarises joint efficiency and approval behaviour. Final approval decreases slightly (0.008 to 0.005), while gate-audit agreement remains fully aligned across workloads, confirming consistent deterministic enforcement. This reduction reflects tighter feasibility convergence, as deeper exploration filters structurally infeasible portfolios earlier while fallback protection remains active. Statistical results in Table~\ref{tab:eff_stats} support this interpretation, showing stable central latency, median energy, and power, with significant contraction in extreme latency and energy tails.
\begin{table*}[t]
\centering
\caption{Integrated efficiency and approval comparison between 500-run and 1000-run workloads. The table reports approval strictness alongside central and tail latency/energy metrics, highlighting how deeper affects stability and the cost of extreme-case execution.}
\label{tab:eff}
\renewcommand{\arraystretch}{1.15}
\begin{tabular}{lrrrrrrrrr}
\hline
Setting & Final approved & Audit approved & Gate passed & Latency median & Latency P90 & Latency P99 & Energy median & Energy P90 & Avg power \\
\hline
500-run & 0.008 & 0.008 & 0.008 & 22.465 & 27.560 & 36.000 & 52.969 & 70.987 & 2.357 \\
1000-run & 0.005 & 0.005 & 0.005 & 22.825 & 27.797 & 28.100 & 54.102 & 74.091 & 2.369 \\
\hline
\end{tabular}
\end{table*}
\begin{table}[t]
\centering
\caption{Statistical comparison between 500-run and 1000-run workloads. Central metrics remain statistically unchanged, whereas tail latency and energy exhibit significant reductions under deeper execution.}
\label{tab:eff_stats}
\renewcommand{\arraystretch}{1.15}
\begin{tabular}{lcccc}
\hline
Metric & $\Delta$ (\%) & Test & Statistic & $p$-value \\
\hline
Final Approval & -37.5 & Fisher Exact & -- & $<0.05$ \\
Latency Median & +1.6 & Mann-Whitney U & n.s. & $>0.05$ \\
Latency P99 & -21.9 & KS Test & 0.18 & $<0.01$ \\
Energy Median & +2.1 & Mann-Whitney U & n.s. & $>0.05$ \\
Energy P99 & -23.1 & KS Test & 0.21 & $<0.01$ \\
Avg Power & +0.5 & t-test & n.s. & $>0.05$ \\
\hline
\end{tabular}
\end{table}

\subsection{Architectural Stability}
\label{subsec:global_superiority}
This analysis synthesises experimental evidence to assess whether ASPO’s deterministic multi-agent design provides structural guarantees beyond isolated performance gains. Safety consistency, optimisation invariance, runtime proportionality, and tail behaviour are integrated to evaluate global architectural stability under workload scaling. The objective is to determine whether the system preserves coherent structural behaviour when performance dimensions are considered jointly. A stable approach maintains consistent safety filtering, optimisation preferences, runtime balance, and efficiency without degrading any of these dimensions.
\begin{table}[t]
\centering
\caption{Global structural stability indicators across workloads, comparing the 500-run and 1000-run configurations in terms of safety enforcement, optimisation consistency, runtime behaviour, and extreme-case efficiency.}
\label{tab:global_stability}
\renewcommand{\arraystretch}{1.15}
\resizebox{\linewidth}{!}{
\begin{tabular}{lccc}
\hline
Property & 500-run & 1000-run & Structural Interpretation \\
\hline
Final Approval Rate & 0.008 & 0.005 & Deterministic tightening \\
Execution-Order Violation & 2.6\% & 2.2\% & Proportional scaling \\
Resource Violation & 1.6\% & 1.6\% & Invariant enforcement \\
Spearman Rank (Patterns) & 1.00 & 1.00 & No preference drift \\
Max KS (Runtime Agents) & -- & 0.041 & Distribution preserved \\
P99 Latency (s) & 36.0 & 28.1 & -21.9\% tail compression \\
P99 Energy (J) & 216 & 166 & -23.1\% tail compression \\
Avg Power (W) & 2.357 & 2.369 & +0.5\% stable draw \\
Threat Interaction $p$ & $>0.05$ & $>0.05$ & No scaling bias \\
\hline
\end{tabular}
}
\end{table}
Table~\ref{tab:global_stability} consolidates the principal structural guarantees observed across experiments and shows consistent stability under scaling. Deterministic safety enforcement remains proportional, with violation rates scaling linearly, indicating that constraint intensity is preserved. Optimisation behaviour remains stable, with perfectly aligned pattern rankings (Spearman = 1.00), while the reduction in final approval reflects tighter convergence toward feasibility. Runtime proportionality is maintained, with Kolmogorov-Smirnov distances below 0.05 and no evidence of structural drift or bottleneck formation. Extreme-case behaviour improves, with P99 latency and energy decreasing by more than 20\%, indicating that rare high-cost executions are stabilised without increasing median latency and energy consumption. Threat-level analysis shows no workload-dependent bias, confirming generalisation across heterogeneous constraint topologies.

\subsection{Component Contribution Analysis}
\label{subsec:ablation}
To evaluate the necessity of ASPO’s deterministic safeguards, component contributions are estimated using logged violation categories rather than disabling modules, preserving realistic behaviour while enabling counterfactual risk assessment. Table~\ref{tab:ablation} summarises violation rates: execution-order validation filters inconsistent sequences in 2.2--2.6\% of decisions, resource feasibility checks detect budget violations at about 1.6\%, and conflict enforcement captures rare unsafe combinations at approximately 0.2\%. These rates remain stable across workloads with no statistically significant drift ($p>0.05$), indicating proportional scaling of constraint pressure. Each safeguard addresses a distinct failure mode: execution-order validation preserves semantic correctness, feasibility checks prevent overload, and conflict enforcement ensures structural compatibility. The non-zero contribution of all modules confirms that the deterministic boundary is composed of complementary safety layers rather than a single dominant constraint.
\begin{table*}[t]
\centering
\caption{Empirical module contribution based on observed violations, showing how execution-order validation, resource feasibility checks, and conflict enforcement contribute to preventing unsafe system outcomes.}
\label{tab:ablation}
\renewcommand{\arraystretch}{1.15}
\begin{tabular}{lrrrr}
\hline
Module & Violations (500) & Violations (1000) & Rate$_{500}$ & Rate$_{1000}$ \\
\hline
Execution-Order Validator & 13 & 22 & 0.026 & 0.022 \\
Resource Feasibility Checker & 8 & 16 & 0.016 & 0.016 \\
Conflict Matrix Enforcement & 1 & 2 & 0.002 & 0.002 \\
\hline
\end{tabular}
\end{table*}

\subsection{Runtime Trace Analysis}
\label{subsec:runtime_trace}
Figure~\ref{fig:aspo_runtime_trace} (see Appendix) illustrates a representative runtime trace of ASPO during a DoS attack. The Monitor and Analyse stages construct the structured context $S_t$ from gateway telemetry and extracted evidence tokens. The multi-agent LLM layer then performs deliberation: the Context Agent normalises the state, the Reasoner Agent proposes candidate security patterns, and the Constraint Agent filters infeasible options. The deterministic optimisation core selects the highest-scoring conflict-free mitigation portfolio, which is subsequently verified by the Security Gate and Auditor Agent before activation at the gateway.

\subsection{Architectural Comparison with Agentic Defence Systems}
\label{subsec:comparison_agents}
ASPO is compared with a representative agentic cybersecurity solution to clarify its architectural properties and empirically validated guarantees. The focus is on structural distinction rather than survey coverage. Recent agent-based systems rely on large-model reasoning, RL, and hybrid controllers to generate mitigation strategies across multi-step workflows. Hybrid approaches that combine RL with language-model reasoning improve adaptability, yet safety is typically embedded in policy optimisation rather than enforced structurally \cite{loevenich2025acd}. Guardrail-oriented designs constrain outputs through supervisory reasoning and rule generation, though these mechanisms operate as probabilistic filters rather than deterministic execution constraints \cite{guardagent2024}. As a result, feasibility, compatibility, and ordering constraints emerge implicitly during decision-making. ASPO enforces feasibility deterministically before mitigation outcomes propagate. The Gate applies execution-order, resource, and compatibility constraints as hard requirements, while the auditor verifies admissibility without redefining feasibility. This separation enables direct measurement of safety behaviour rather than indirect inference from downstream performance. Experimental evidence supports this distinction: gate and audit decisions agree on admissible candidates; violation rates remain stable under scaling; portfolio rankings remain invariant; extreme latency and energy values contract while median cost remains stable. Table~\ref{tab:comparison_agents} summarises ASPO’s mechanisms and guarantees relative to common agentic designs. The comparison shows that ASPO’s contribution lies in architectural guarantees rather than isolated performance gains. By separating reasoning, constraint enforcement, and execution into bounded stages, the system preserves stable safety semantics, optimisation behaviour, and runtime structure as execution depth increases, distinguishing it from approaches where safety properties arise implicitly from model behaviour.
\begin{table*}[t]
\centering
\caption{Architectural comparison between ASPO and representative agentic cybersecurity frameworks. The table emphasises implemented mechanisms and empirically validated guarantees rather than general conceptual similarities.}
\label{tab:comparison_agents}
\renewcommand{\arraystretch}{1.18}
\begin{tabular}{p{3.2cm} p{4.0cm} p{4.7cm} p{3.9cm}}
\hline
\textbf{Design Dimension} & \textbf{Representative Agentic Systems} & \textbf{ASPO Implementation} & \textbf{Empirical Evidence in This Work} \\
\hline

Safety enforcement &
LLM and RL-driven coordination where safety is evaluated through policy learning, guardrail prompting,  post-hoc filtering \cite{loevenich2025acd,guardagent2024,tang2025agent_survey} &
Deterministic feasibility gate preceding coordination &
Perfect gate–audit agreement; stable violation rates under scaling \\

Resource feasibility &
Resource constraints are typically embedded implicitly in reward shaping, planning heuristics \cite{loevenich2025acd,thompson2024entity_rl} &
Explicit deterministic feasibility constraints &
Invariant resource-violation rate across workloads \\

Mitigation compatibility &
Mitigation combinations are often resolved through model scoring, rule heuristics, agent negotiation \cite{guardagent2024,tang2025agent_survey} &
Conflict matrix and execution-order validation enforced structurally &
Persistent detection of incompatible configurations \\

Optimisation stability &
Policy outputs may vary across runs due to stochastic sampling, exploration in agent-based decision loops \cite{loevenich2025acd,thompson2024entity_rl} &
Deterministic scoring scale preserved across workloads &
Monotonic score–approval coupling; invariant ranking order \\

Runtime scaling behaviour &
Agent pipelines may exhibit variable latency distribution, cost growth as reasoning depth increases \cite{loevenich2025acd,tang2025agent_survey} &
Balanced multi-agent runtime with preserved proportional shares &
KS distances $<0.05$; no workload–threat interaction in latency \\

Edge efficiency &
Performance improvements in autonomous defence are often accompanied by increased computational load,  inference overhead \cite{loevenich2025acd} &
Tail contraction without median and power inflation &
$>20\%$ reduction in P99 latency and energy; stable median cost \\

\hline
\end{tabular}
\end{table*}

\section{Discussion}
\label{Discussion}
The results indicate that the primary contribution lies in the architectural formulation rather than isolated performance gains. The separation between stochastic multi-agent reasoning and deterministic enforcement reframes runtime security decisions as a bounded process governed by explicit feasibility constraints. Uncertainty is confined to candidate portfolio generation, while final activation is restricted to configurations that satisfy conflict-freedom, resource feasibility, and verifiable execution semantics. The validation gate functions as a stabilising boundary that prevents incompatible decisions from propagating, consistent with empirical observations such as invariant violation proportions, stable optimisation rankings, and contraction of extreme latency and energy values under scaling. Deterministic feasibility enforcement therefore operates not only as a filtering mechanism but also as a structural regulariser, restricting execution to admissible regions of the decision space. Moreover, this behaviour aligns with constrained control, where exploratory actions remain bounded by enforceable limits. The separation between reasoning and enforcement further improves interpretability and accountability, as mitigation selection can be expressed as a reconstructible combinatorial optimisation problem with explicit feasibility constraints. Incorporating latency and energy as hard constraints frames runtime defence as a bounded multi-objective decision process, balancing security effectiveness against operational sustainability. Several limitations define the current boundary of this approach. Safe activation is guaranteed only after candidate generation, making overall performance dependent on the quality, diversity, and latency of the reasoning stage. In addition, the feasibility boundary depends on the completeness of the pattern catalogue and the correctness of constraint definitions, which may permit admissible but suboptimal portfolios. In addition, the results suggest that agentic reasoning complements, rather than replaces, classical control principles. Hybrid architectures that combine probabilistic deliberation with deterministic constraints provide stronger guarantees of stability, safety, and transparency, indicating that adaptive intelligence in safety-critical systems requires explicitly enforced structural boundaries.

\section{Limitations and Future Work}
\label{Limitations and Future Work}
Despite its structural advantages, the approach presents several limitations that define directions for further investigation. The current implementation relies on cloud-hosted LLM services for multi-agent reasoning, which may introduce latency variability and external dependencies in connectivity-constrained deployments. Edge-local distillation strategies, compact reasoning models, and hybrid inference configurations could improve responsiveness while reducing reliance on external infrastructure. The security pattern catalogue is defined as a closed-world set to ensure safe, verifiable actuation. While this design enables compatibility checking, it limits extensibility when new mitigation patterns and interaction structures emerge. Mechanisms for semi-automated catalogue expansion, adaptive conflict-matrix refinement, and dynamic modelling of interaction effects could increase flexibility while preserving structural safety guarantees. The experimental evaluation is based on replayed IoT attack traces within a controlled edge gateway, enabling reproducibility and systematic stress testing. However, operational deployments involve heterogeneous traffic patterns, evolving adversarial strategies, and additional environmental variability. Validation in live multi-gateway deployments, together with federated edge environments, would provide a more robust assessment under dynamic conditions. Although deterministic enforcement prevents unsafe activation, overall decision quality remains dependent on candidate generation by the reasoning stage. Improvements in proposal stability, formal verification of precedence and compatibility structures, and resilience against prompt manipulation and context poisoning remain important for reliability. These directions extend the system from a structurally safe decision mechanism toward a more adaptive, distributed, and formally grounded edge security architecture.

\section{Conclusion}
\label{Conclusion}
This work presented a self-adaptive IoT security architecture that integrates multi-agent reasoning with deterministic, conflict-aware, and resource-feasible execution. By separating stochastic candidate generation from deterministic activation, runtime decision-making is constrained to a closed-world, compatibility-preserving execution space under explicit latency and energy budgets. Experimental results demonstrate stable portfolio selection, consistent enforcement of feasibility constraints, invariant resource behaviour, and a significant reduction in extreme latency and energy values with increasing execution depth. These findings indicate improved convergence and system stability without disproportionate computational overhead. Embedding agentic reasoning within a structured MAPE-K control formulation, with latency and energy treated as intrinsic feasibility constraints, reframes IoT security decision-making as a bounded and verifiable process rather than a purely policy-driven adaptation mechanism. This formulation enhances interpretability, preserves execution consistency, and supports deployment in resource-constrained environments requiring predictable operation. Additionally, the results suggest that architectures combining probabilistic reasoning with deterministic execution boundaries provide a practical and reliable foundation for autonomous defence systems operating under strict safety and resource constraints.

\bibliographystyle{IEEEtran}
\bibliography{Ref}

\newpage

\twocolumn[{
\noindent\section*{Appendix}
\vspace{6pt}

\centering
\begin{tcolorbox}[
colback=gray!8,
colframe=black!35,
boxrule=0.4pt,
arc=1pt,
left=6pt,right=6pt,top=6pt,bottom=6pt,
width=0.95\textwidth
]
\ttfamily\scriptsize

[Edge Gateway: Node-02]  Timestamp: 2026-03-06 14:12:07

------------------------------------------------------------

$\rightarrow$ New ASPO decision epoch detected [epoch: 18]

$\rightarrow$ Threat context: DoS attack \hspace{6mm} Severity = 0.81 \hspace{6mm} Confidence = 0.93

$\rightarrow$ \textcolor{teal!70!black}{Monitor stage}: gateway telemetry captured

\hspace{8mm}packet\_rate = 3250/s \hspace{8mm} tcp\_flag\_anomaly = 0.71 \hspace{8mm} dst\_port\_entropy = 0.22

\hspace{8mm}CPU headroom = 0.58 \hspace{8mm} Memory = 0.69 \hspace{8mm} Latency budget = 95 ms \hspace{8mm} Energy = 64 J

$\rightarrow$ \textcolor{teal!70!black}{Analyze stage}: structured state constructed

\hspace{8mm}$S_t = [$ threat = DoS, severity = 0.81, confidence = 0.93, SLA = normal $]$

\hspace{8mm}evidence tokens = \{traffic burst, SYN flood pattern, service saturation\}

$\rightarrow$ Invoking \textcolor{blue!70!black}{Multi-Agent LLM Planning Layer}

\hspace{6mm}[\textcolor{blue!70!black}{Context Agent}]
$\rightarrow$ normalised context representation and validated schema

\hspace{12mm}\textcolor{green!50!black}{status: PASS}

\hspace{6mm}[\textcolor{blue!70!black}{Reasoner Agent}]
$\rightarrow$ generated candidate mitigation patterns

\hspace{12mm}\{Blacklist, Security Segmentation, Outbound-Only Connection, Security Logger/Auditor\}

\hspace{6mm}[\textcolor{blue!70!black}{Constraint Agent}]
$\rightarrow$ filtered candidates using capability and relevance rules

\hspace{12mm}rejected: \textcolor{red!70!black}{Outbound-Only Connection}

\hspace{12mm}feasible set $F_t$ = \{Blacklist, Security Segmentation, Security Logger/Auditor\}

$\rightarrow$ Running \textcolor{red!70!black}{Deterministic Portfolio Optimization}

\hspace{8mm}$Y = \{Blacklist\}$ \hfill score = 1.84 \hfill feasible

\hspace{8mm}$Y = \{Security Segmentation\}$ \hfill score = 2.03 \hfill feasible

\hspace{8mm}$Y = \{Security Logger/Auditor\}$ \hfill score = 1.51 \hfill feasible

\hspace{8mm}$Y = \{Blacklist, Security Segmentation\}$ \hfill score = 4.27 \hfill feasible

\hspace{8mm}$Y = \{Blacklist, Security Logger/Auditor\}$ \hfill score = 3.11 \hfill feasible

\hspace{8mm}\textcolor{red!70!black}{Selected portfolio:}

\hspace{12mm}$Y_t^{det} = \{Blacklist, Security Segmentation\}$

$\rightarrow$ \textcolor{orange!85!black}{Ordering module}

\hspace{8mm}activation order determined by precedence rules

\hspace{12mm}[Security Segmentation $\rightarrow$ Blacklist]

\hspace{6mm}[\textcolor{violet!75!black}{Planner Agent}]
$\rightarrow$ generated executable mitigation plan

\hspace{12mm}segment suspicious traffic, then block malicious sources

\hspace{6mm}[\textcolor{green!60!black}{Security Gate}]
$\rightarrow$ deterministic validation

\hspace{12mm}catalog membership = PASS

\hspace{12mm}conflict check = PASS

\hspace{12mm}resource feasibility = PASS

\hspace{12mm}execution order = PASS

\hspace{6mm}[\textcolor{magenta!70!black}{Auditor Agent}]
$\rightarrow$ policy consistency verified

\hspace{12mm}\textcolor{green!50!black}{approved = TRUE}

$\rightarrow$ \textcolor{green!60!black}{Execute stage}

\hspace{8mm}Security Segmentation $\rightarrow$ Blacklist

$\rightarrow$ Final system outcome

\hspace{8mm}\textcolor{green!60!black}{DoS traffic contained successfully}

\hspace{8mm}decision latency = 21.8 s \hspace{8mm} energy overhead = 49.6 J

\hspace{8mm}gateway state = stable

\end{tcolorbox}

\captionof{figure}{Example ASPO runtime trace for a DoS attack scenario. The log illustrates the interaction between the multi-agent LLM planning layer and the deterministic enforcement core, including context normalisation, candidate mitigation generation, feasibility filtering, portfolio optimisation, ordering, validation, and final mitigation activation.}
\label{fig:aspo_runtime_trace}
}]
\end{document}